\pdfoutput=1
\documentclass[12pt]{article}
\usepackage[margin=3cm]{geometry}
\usepackage[utf8]{inputenc}
\usepackage[T1]{fontenc}
\usepackage[english]{babel}
\usepackage{times}
\usepackage{setspace}
\usepackage{authblk}
\usepackage{graphicx}
\usepackage{booktabs}
\usepackage{amsmath}
\usepackage{bm}
\usepackage{tikz}
\usetikzlibrary{positioning,decorations.pathreplacing}
\usepackage{adjustbox}
\definecolor{cgreen}{HTML}{33A02C}
\definecolor{cblue}{HTML}{1F78B4}
\usepackage[authoryear]{natbib}
\usepackage[hidelinks]{hyperref}
\title{\textbf{Anticipating Continued Global Fertility Decline via Neural Forecasting}}
\ifdefined\anonymous
\author{}
\hypersetup{
  pdftitle={Anticipating Continued Global Fertility Decline via Neural Forecasting}
}
\else
\author[1,2]{Daniel Ciganda\thanks{\href{mailto:ciganda@demogr.mpg.de}{ciganda@demogr.mpg.de}}}
\author[2]{Facundo Morini}
\author[3]{Francisco Piriz}
\author[1,5]{\mbox{Henrik-Alexander Schubert}}
\author[1]{Ugofilippo Basellini}
\author[1,4,5]{\mbox{Mikko Myrskyl\"a}}

\affil[1]{Max Planck Institute for Demographic Research, Rostock, Germany}
\affil[2]{Statistics Institute, University of the Republic, Montevideo, Uruguay}
\affil[3]{Digital Sense, Montevideo, Uruguay}
\affil[4]{Helsinki Institute for Demography and Population Health, University of Helsinki, Helsinki, Finland}
\affil[5]{Max Planck -- University of Helsinki Center for Social Inequalities in Population Health, Rostock, Germany; Helsinki, Finland}
\hypersetup{
  pdftitle={Anticipating Continued Global Fertility Decline via Neural Forecasting},
  pdfauthor={Daniel Ciganda; Facundo Morini; Francisco Piriz; Henrik-Alexander Schubert; Ugofilippo Basellini; Mikko Myrskyl\"a}
}
\fi
\date{}

\begin{document}
\maketitle

\begin{abstract}
As fertility falls across much of the world, major projection systems disagree about whether decline will give way to recovery, reflecting different assumptions about post-transitional fertility. We ask instead what the historical record suggests, using NeuralTFR, a recurrent neural network that learns fertility dynamics solely from pooled histories of 230 countries and territories. In out-of-sample forecasts from three origins, it predicts fifteen years ahead as accurately as the United Nations' Bayesian model, and its prediction intervals score slightly better. Its forecasts reveal a simple nonlinear relationship, which a spline regression reproduces with similar accuracy: fertility tends to rise slightly from the lowest levels and decline from higher ones, with recent trends affecting the size of these changes more than their direction. By 2038, these dynamics leave very low fertility, below 1.5 births per woman, more widespread than the United Nations expects, without the extreme lows of the most pessimistic projections.

\medskip
\textit{Keywords:} Demographic forecasting; Population projections; Probabilistic forecasting; Recurrent neural networks; Quantile regression; Cross-learning; Interpretability
\end{abstract}

\section{Introduction}
\label{sec:introduction}

The accelerating global transition toward low and very low fertility poses profound challenges to economic growth, welfare systems, and geopolitical stability. Anticipating the extent and duration of this demographic shift requires reliable projections of the total fertility rate (TFR). A central debate in demographic forecasting concerns whether recent fertility declines will be followed by stabilization and recovery, how far that recovery might extend, and whether very low fertility will become more widespread \citep[see, e.g.,][]{basten2014future}.

Global demographic forecasting is currently shaped by three major institutional approaches: the United Nations Population Division \citep{un2024wpp}, the Wittgenstein Centre for Demography and Global Human Capital \citep{kc2024updating}, and the Global Burden of Disease (GBD) project \citep{gbd2024global}. These approaches differ in both their methods and projected outcomes, reflecting different assumptions about post-transitional fertility, the role of expert judgment, and the importance of external factors. Underlying these differences is a gap in demographic knowledge: post-transitional fertility has been observed in only a handful of countries over a limited period \citep{goujon2025}, leaving limited empirical evidence to guide long-run projections.

The expectation of recovery draws mainly on the European part of this limited experience. During the 1990s, fertility fell below 1.3 births per woman in much of Southern, Central, and Eastern Europe, a level described as lowest-low fertility \citep{kohler2002emergence}, and then rose in most of these countries during the following decade \citep{goldstein2009end}. Much of this rise reflected a slowdown in the postponement of childbearing, which had kept period fertility below the number of children women would eventually have \citep{bongaarts2012demographic}. The same experience underlies the recovery in the United Nations projections: the post-transition phase of BayesTFR was first estimated from 21 countries in which fertility had risen after the transition, 19 of them European \citep{raftery2014bayesian}.

Most of these recoveries, however, did not last. In the data assembled for this study (Section~\ref{sec:data}), 55 countries and territories, including China, Japan, Russia, and most of Europe, had fertility below 1.5 births per woman by 2006. In 40 of them, fertility later rose by at least 0.2 births per woman from its lowest level, but 32 of these have since fallen back below 1.5. More than half of the 55 now have fertility at or below the lowest level they had reached by 2006. The historical record therefore offers little support for a lasting recovery process. This raises the question of what recent fertility trajectories suggest when projections rely less on assumptions about recovery, external covariates, and expert adjustment.

We address this question by developing NeuralTFR, a recurrent neural network that learns fertility change from harmonized histories of 230 countries and territories. The model generates central forecasts and prediction intervals from countries' past fertility trajectories, without socioeconomic covariates, prespecified transition phases, or long-run recovery levels. In fifteen-year forecasts from three historical origins (1995, 2000, and 2006), NeuralTFR improves on a drift benchmark and matches the point accuracy of BayesTFR, the Bayesian hierarchical fertility model underlying the United Nations projections, while its 80\% and 95\% prediction intervals have slightly lower Winkler scores. NeuralTFR's point forecasts are more accurate where fertility is already very low, the range in which recovery is most often assumed and least often observed.

Since no form of fertility change is imposed, understanding what NeuralTFR learns is informative in itself. The analysis of its forecasting patterns reveals a simple, nonlinear map from the current fertility level to subsequent change, modified by the recent trajectory. We then express this map explicitly in a pooled spline quantile regression, using the flexible model as a guide to the relationships a transparent model should represent \citep{craven1996extracting,rudin2019stop}. The regression relates future fertility to its current level and recent changes, allowing these relationships to change slope across their range and to evolve smoothly across forecast years. Estimated from the same country histories, this spline model produces forecasts and prediction intervals of broadly similar quality to NeuralTFR's.

We use this learned map to provide a data-driven assessment of global fertility trajectories over a 15-year horizon. In a comparison covering 130 countries and territories, NeuralTFR and the spline model both project more populations below 1.5 births per woman than the United Nations, but fewer below 1.3 than GBD. Our models also project some recovery, because fertility has often risen within fifteen years of reaching its lowest levels. What distinguishes them from the United Nations is the level at which this recovery gives way to decline. Both NeuralTFR and the spline model place it close to its observed level, at about 1.5 births per woman, whereas BayesTFR places it higher and predicts larger increases below it. In our projections, fertility therefore rises only modestly where it is already lowest and continues to fall in populations just above 1.5, so that very low fertility becomes more widespread by 2038. What emerges from the historical record, then, is a future in which very low fertility becomes more common rather than much lower, although the modest recovery we project at the lowest levels may itself prove optimistic, given how few past recoveries have lasted.

\subsection{Previous Work}

The three institutional systems introduced above differ most in how they specify post-transitional fertility and its long-run evolution.

The first, and most widely adopted, is the probabilistic framework used by the United Nations Population Division \citep{un2024wpp}, which relies on the Bayesian Hierarchical Model (BayesTFR) developed by Raftery, \v{S}ev\v{c}{\'i}kov{\'a}, Alkema and colleagues \citep{alkema2011probabilistic,raftery2014bayesian,raftery2023probabilistic}. BayesTFR combines coherent probabilistic projections with a demographic model that distinguishes phases of the fertility transition. Its extensive use in official projections makes it a central reference for this comparison. In the post-transition phase (Phase III), an autoregressive AR(1) process draws fertility toward a country-specific long-term level. That level depends in part on a global prior derived from countries where fertility has already recovered. Stabilization and recovery toward the long-term level are thus built into the model. For instance, the UN explicitly assumes that fertility in countries with already-peaked populations will gradually rebound to an average of 1.4 births per woman by 2100 \citep{un2024wpp}. Reflecting this expectation of recovering fertility, the UN anticipates the most conservative fertility decline of the three systems, forecasting a global TFR of about 1.8 births per woman and a global population of 10.2 billion by 2100 \citep{un2024wpp,un2025fertility}.

The second approach, developed by the Wittgenstein Centre for Demography and Global Human Capital (IIASA/VID), uses multi-state cohort-component models that distinguish populations by educational attainment \citep{lutz2014world,lutz2018demographic,kc2024updating}. Panels of demographic experts determine the long-term fertility levels, rather than an automated extrapolation of past time series. In the most recent assessment (WIC2023), the projections remain close to those of the United Nations over much of their common horizon. The medium scenario gives a global TFR of about 1.7 births per woman in 2095--2100 and a global population of about 9.9 billion in 2100 \citep{kc2024updating,un2025fertility}.

A third approach was developed by the Global Burden of Disease (GBD) project at the Institute for Health Metrics and Evaluation (IHME). The 2024 GBD model projects fertility using an ensemble approach with four covariates: female educational attainment, contraceptive met need, under-5 mortality, and population density in habitable areas \citep{gbd2024global}. Linking future fertility to these projected drivers, the GBD model anticipates a steeper and more sustained fertility decline than the UN. Its reference scenario projects a global TFR of 1.59 births per woman by 2100, the lowest of the three systems \citep{gbd2024global,un2025fertility}. The resulting projections depend on how these drivers are expected to evolve over the coming decades, and on fertility continuing to respond to them as it has in the past. Earlier IHME projections were criticized for the uncertainty of these assumptions, as well as for the quality of their baseline data and their treatment of post-transitional fertility \citep{gietel2020uncertain,gietel2021trends}.

A central challenge in fertility forecasting is that individual country series are short and heterogeneous, while the broader panel contains shared temporal structure. Demographic forecasting has a long tradition of learning such structure statistically. Principal-component methods use past rates to estimate age-specific patterns of change \citep{lee1992modeling,hyndman2007robust}. Penalized splines estimate smooth patterns over age and time, with the smoothing itself providing the basis for forecasts \citep{currie2004smoothing}. Constraining these projections to age profiles and trends derived from observed data offers a middle ground between purely statistical and traditional demographic models (\citealp{camarda2019smooth,camarda2021smoothing}; see \citealp{basellini2023thirty}, for a review).

Similar ideas have been developed for completed cohort fertility. \citet{schmertmann2014bayesian} derive priors from historical fertility schedules pooled across the Human Fertility Database, so that forecasts favor age patterns and trends resembling those observed in the past. \citet{ellison2020forecasting} forecast cohort fertility for 30 countries with a hierarchical Bayesian model, achieving accuracy comparable with the best-performing existing methods, particularly for countries without a recent structural shift, and later extend this work to parity \citep{ellison2026parity}. Across populations, information is more often shared within a prespecified form of change, as in coherent mortality forecasts \citep{li2005coherent} and BayesTFR's hierarchical transition model \citep{alkema2011probabilistic}. A development closer to our approach is that of \citet{susmann2025flexible}, who model only the decline from high to low fertility (Phase~II), replacing BayesTFR's parametric function with B-splines estimated across countries. In out-of-sample checks, their projections generally had lower errors than those based on the parametric function, showing that learning the form of decline from the data can improve projections.

The forecasting literature has emphasized the value of such sharing. Models estimated jointly across many series can be simple yet competitive \citep{montero2021principles}, and learning across series underlies much recent progress in forecasting, from forecasting competitions \citep{makridakis2022m5} to foundation models pretrained on large collections of time series \citep{ansari2024chronos}. Because pooling and flexible architectures usually arrive together, their contributions are easily confounded. Comparing recurrent networks with pooled regressions, \citet{hewamalage2021recurrent} find that the architecture adds accuracy beyond pooling on most of their datasets, whereas in mortality forecasting most of the gain comes from pooling populations \citep{park2025next}.

\section{Data and methods}
\label{sec:data_methods}

\subsection{Data}
\label{sec:data}

We construct annual fertility trajectories from 55,194 TFR estimates covering 230 countries and territories between 1950 and 2025. Most estimates (96\%) come from the compilation of empirical fertility estimates maintained by the United Nations Population Division. These empirical source estimates are distinct from the modeled historical series published in World Population Prospects. Using the source estimates allows us to learn from variation in reported fertility and evaluate forecasts without inheriting the additional assumptions used to reconstruct historical trajectories.

We updated this compilation through a country-by-country review of official statistical websites, adding recent fertility estimates from national statistical offices, Eurostat and the Economic Commission for Latin America and the Caribbean (ECLAC). These additions provide the most recent observation for 178 of the 230 populations. The data, including the source of each estimate, are available in the study's repository (\url{https://github.com/FaMori/NeuralTFR}). The estimates are derived from censuses (44\%), surveys (25\%), vital and population registers (13\%), and other published estimates and official statistics (18\%). About two-thirds are indirect estimates, obtained by applying demographic methods to census and survey information rather than calculating fertility directly from reported births. Where multiple estimates are available for the same country-year, we take their median to construct the annual series.

Interior gaps in the annual series are filled by linear interpolation, accounting for 3.2\% of country-year values in the completed panel. We take natural logarithms of TFR and center and scale the resulting values using the same fixed constants for all populations. Forecast errors are calculated only for country-years with estimates in the original data, excluding interpolated values.

\subsection{Study design}
\label{sec:design}

We compare the performance of different models using a rolling-origin design with three forecast origins: 1995, 2000, and 2006. At each origin, models are trained using fertility observations up to that year. We evaluate NeuralTFR's fifteen-year forecasts against BayesTFR and the spline regression introduced above. We also evaluate a drift model and a linear version of the pooled regression to assess whether the more flexible models improve on simple trend extrapolation and linear relationships between predictors and future fertility.

For the forward projections, we fit NeuralTFR and the spline model using observations through 2023 and project fifteen years ahead, to 2038. We compare these forecasts with published United Nations and GBD projections.

\begin{samepage}
Model performance may vary across fertility regimes, so aggregate results can conceal differences in how well models anticipate continued decline, stabilization, or recovery. We therefore examine model performance and forward projections separately for three groups, defined by the TFR at the forecast origin: very low fertility ($\mathrm{TFR} < 1.5$), moderately low fertility ($1.5 \leq \mathrm{TFR} < 2.1$), and fertility at or above replacement ($\mathrm{TFR} \geq 2.1$). We use ``low fertility'' for the broader category below 2.1 births per woman. Figure~\ref{fig:coverage_composition} in Appendix~\ref{app:fertility_composition} shows how the representation of these groups in the panel changes over time. Because very low fertility spread mainly in recent decades, the historical origins offer limited experience of it: up to 1995, 2000, and 2006, between 34 and 55 populations had reported very low fertility, accounting for 3--5\% of reported population-years, compared with 89 populations and 8\% up to 2023. The historical evaluation is therefore based on less experience of very low fertility than the forward projections.
\par\end{samepage}

\subsection{The NeuralTFR model}
\label{sec:neuraltfr}

We convert each country's fertility series into input--output pairs using a sliding-window approach. Each pair consists of an input sequence covering twelve consecutive years of fertility history and a target sequence containing subsequent fertility values. The window shifts forward one year at a time through each country's series, generating overlapping pairs that capture fertility trajectories beginning at different points in time. These pairs are pooled across countries to train a single global model with shared parameters \citep{januschowski2020criteria}. At each pass through the training data, these pooled windows are shuffled and each window is used once.

Let $z_{i,t}=(\log \mathrm{TFR}_{i,t}-a)/b$ denote the transformed fertility level for country $i$ in year $t$, where $a$ and $b$ are fixed constants, shared across countries and origins, that only put the inputs on a scale suited to training. Because the transformation is linear, their exact values do not change the information the inputs carry (Appendix~\ref{app:pooled_regressions}). For a window ending in year $t$, the input in each year $s=t-11,\ldots,t$ is $\boldsymbol{x}_{i,s}=(z_{i,s},d_{i,s})$, where $d_{i,s}=(z_{i,s}-z_{i,s-1})/s_\Delta$ is the year-to-year difference in transformed fertility divided by its standard deviation across the training windows, $s_\Delta$. We set the difference at the first position of each window to zero. The input history and target sequence are
\begin{equation}
X_{i,t}=(\boldsymbol{x}_{i,t-11},\ldots,\boldsymbol{x}_{i,t}),
\qquad
Y_{i,t}=(z_{i,t+1},\ldots,z_{i,t+H}),
\label{eq:training_windows}
\end{equation}
where $H$ is the maximum target length: 15 years in both the historical evaluation and the forward projections. We also include training windows for which fewer than fifteen subsequent years are available, calculating the loss only for reported values within the training period. Figure~\ref{fig:sliding_windows} illustrates the sliding-window construction.

\begin{figure}[!htbp]
\centering
\begin{adjustbox}{max width=\linewidth}
\begin{tikzpicture}[
	inputbox/.style={
		draw,
		rectangle,
		draw=cblue!50,
		fill=cblue!30,
		minimum width=1.4cm,
		minimum height=1.2cm,
		align=center,
		font=\scriptsize\bfseries
	},
	targetbox/.style={
		draw,
		rectangle,
		draw=cgreen!50,
		fill=cgreen!30,
		minimum width=1.4cm,
		minimum height=1.2cm,
		align=center,
		font=\scriptsize\bfseries
	},
	brace/.style={
		decorate,
		decoration={brace, amplitude=6pt, raise=4pt},
		line width=0.8pt
	},
	bracelabel/.style={
		font=\small,
		above=10pt
	}
	]

	\begin{scope}[local bounding box=window1]
		\node[inputbox] (t0) {$\bm{t-11}$};
		\node[inputbox, right=0 of t0] (t1) {$\bm{t-10}$};
		\node[inputbox, right=0 of t1] (dots1a) {\dots};
		\node[inputbox, right=0 of dots1a] (t22) {$\bm{t-1}$};
		\node[inputbox, right=0 of t22] (t23) {$\bm{t}$};

		\node[targetbox, right=0 of t23] (t24) {$\bm{t+1}$};
		\node[targetbox, right=0 of t24] (t25) {$\bm{t+2}$};
		\node[targetbox, right=0 of t25] (dots1b) {\dots};
		\node[targetbox, right=0 of dots1b] (t46) {$\bm{t+H-1}$};
		\node[targetbox, right=0 of t46] (t47) {$\bm{t+H}$};

		\draw[brace] (t0.north west) -- (t23.north east)
		node[midway, bracelabel] {Input};
		\draw[brace] (t24.north west) -- (t47.north east)
		node[midway, bracelabel] {Target};

		\draw[brace] (t47.south east) -- (t0.south west)
		node[midway, below=10pt, font=\small] {Window \#1};
	\end{scope}

	\begin{scope}[yshift=-3.5cm, local bounding box=window2]
		\node[inputbox] (t0) {$\bm{t-10}$};
		\node[inputbox, right=0 of t0] (t1) {$\bm{t-9}$};
		\node[inputbox, right=0 of t1] (dots1a) {\dots};
		\node[inputbox, right=0 of dots1a] (t22) {$\bm{t}$};
		\node[inputbox, right=0 of t22] (t23) {$\bm{t+1}$};

		\node[targetbox, right=0 of t23] (t24) {$\bm{t+2}$};
		\node[targetbox, right=0 of t24] (t25) {$\bm{t+3}$};
		\node[targetbox, right=0 of t25] (dots1b) {\dots};
		\node[targetbox, right=0 of dots1b] (t46) {$\bm{t+H}$};
		\node[targetbox, right=0 of t46] (t47) {$\bm{t+H+1}$};

		\draw[brace] (t0.north west) -- (t23.north east)
		node[midway, bracelabel] {Input};
		\draw[brace] (t24.north west) -- (t47.north east)
		node[midway, bracelabel] {Target};

		\draw[brace] (t47.south east) -- (t0.south west)
		node[midway, below=10pt, font=\small] {Window \#2};
	\end{scope}

\end{tikzpicture}
\end{adjustbox}
\caption{\textbf{Schematic representation of the sliding-window approach.} Blue boxes represent the twelve-year input sequence, and green boxes the target sequence of the following $H$ years. Both sequences shift forward by one year to form the next training pair.}
\label{fig:sliding_windows}
\end{figure}

The input sequences contain fertility levels and annual changes, calculated on the transformed scale described above; the network receives neither country identifiers nor external covariates. NeuralTFR processes these inputs through an encoder and a decoder, each built from a single gated recurrent unit (GRU) layer with 16 hidden units and separate learned parameters \citep{cho2014learning,sutskever2014sequence}. The encoder reads the twelve-year history sequentially and summarizes it in a learned representation that initializes the decoder. Starting from the final input value, the decoder generates forecasts one year at a time. At each step, it receives the previous predicted median and the forecast step, expressed as a fraction of the horizon, and produces a set of predictive quantiles; the predicted median is fed into the next step in the same way during training and forecasting. This design combines features of two recurrent global forecasters: like DeepAR \citep{salinas2020deepar}, NeuralTFR feeds its own forecasts back into the network, and like MQ-RNN \citep{wen2017multi}, it produces predictive quantiles trained with a quantile loss rather than the parameters of a predictive distribution.

We estimate five quantiles, $\tau\in\mathcal{Q}=\{0.025,0.10,0.50,0.90,0.975\}$, yielding median forecasts and 80\% and 95\% prediction intervals. We train the model by minimizing the mean quantile loss across these five quantiles and the reported target values in each training batch \citep{koenker1978regression}. For a target $z$ and its predicted quantile $\hat z^{(\tau)}$, this loss is
\begin{equation}
\rho_\tau\!\left(z-\hat z^{(\tau)}\right)
=\max\!\left\{\tau\left(z-\hat z^{(\tau)}\right),\,
(\tau-1)\left(z-\hat z^{(\tau)}\right)\right\}.
\label{eq:quantile_loss}
\end{equation}
The loss penalizes underprediction more heavily for upper quantiles and overprediction more heavily for lower quantiles.

To determine how long to train the model, we examine whether additional training improves forecasts of observations withheld from estimation. At each forecast origin, we set aside the ten-year period ending at that origin for validation and fit the model to the earlier observations. Validation forecasts use only inputs observed before this period, so we exclude the few populations whose most recent inputs would otherwise be interpolated toward observations inside it. We select the training duration using mean quantile loss on the original TFR scale, then initialize a new model and train it on all data available up to the origin. Training uses the Adam optimizer with a learning rate of 0.0003, weight decay of 0.0001 and batches of 64 windows. We evaluate the validation loss every ten passes through the training data, up to 200 passes, stop after five evaluations without improvement, and keep the number of passes with the lowest loss. We repeat this procedure ten times, changing the random initialization and the order of the training windows, to reduce dependence on any one fit. For each population and forecast year, we take the median across fits separately for each predictive quantile, which limits the influence of any single fit. These combined quantiles give the central forecast and the bounds of the 80\% and 95\% prediction intervals.

\subsection{Regression models and benchmarks}
\label{sec:benchmarks}

The spline regression is guided by an analysis of how NeuralTFR translates fertility histories into forecasts. We examine how the network's predictions respond to current fertility and the preceding trajectory, passing modified histories through the fitted network without changing its parameters. We first simulate twelve-year TFR histories with different fertility levels and trends. At a given level, this lets us compare forecasts depending on whether fertility has arrived there through decline, stability, or recovery; across levels, it shows how those responses vary. We then assess how much the preceding trajectory contributes to accuracy by replacing each observed history with a constant sequence at its final value and evaluating the resulting forecasts against the same outcomes as those from observed histories. Together, the two exercises point to a simple structure: the network's forecasts depend mainly on current fertility, rising from the lowest levels and declining from higher ones, while the preceding trend alters the size of these changes more than their direction (Appendix~\ref{app:interpretability}).

The spline represents this structure directly, using current fertility and recent annual changes to predict future fertility, with both predictors and outcomes expressed on the transformed scale. For each predictor, we estimate slopes between joining points placed at the deciles of its distribution in the training windows. The resulting connected straight-line segments allow the relationship to change across the predictor's range. Current fertility can therefore relate differently to future fertility at low and high starting levels. We also fit a linear regression using the same predictors as linear terms. Like NeuralTFR, both regressions are estimated from the pooled training windows; for each forecast year, they relate the five predictive quantiles directly to the observed history.

We estimate both models using the quantile loss defined above and select the number of recent annual changes by minimizing validation loss summed across the three historical forecast origins. Both models keep the three most recent changes. We smooth the spline coefficients across forecast years by penalizing abrupt changes in those that determine the median and interval widths. In estimating the relationship for year ten, we therefore draw on years nine and eleven without requiring the three relationships to be identical. We choose the penalty strength using the ten-year validation period ending at each origin. The same value minimizes validation loss at each of the three origins and when we sum their losses. We then refit both models using observations available up to each origin. Appendix~\ref{app:pooled_regressions} provides a more detailed description of the models, estimation process, and sensitivity to the penalty strength.

We also evaluate NeuralTFR against BayesTFR, the established demographic forecasting framework described above. We use the annual implementation of bayesTFR, with autoregression in the fertility-transition phase and historical TFR estimates treated as fixed inputs. For the historical comparisons, we re-estimate both the transition and recovery components at each forecast origin under the package's default prior specifications, using the same data used to train the other models, restricted to observations up to that year. We obtain median forecasts and 80\% and 95\% prediction intervals from its simulated predictive trajectories.

The drift model provides a reference for straightforward extrapolation of recent fertility trends. It serves as a simple, transparent benchmark for assessing whether more complex models improve forecasting performance. It estimates a trend separately for each country on the logarithmic TFR scale, using a period of five to ten years selected from earlier forecast performance. To construct prediction intervals, we make forecasts from earlier points in the observed history and pool their errors across countries, separately for each forecast horizon. The distribution of these errors is used to adjust the central forecast and set the lower and upper interval bounds.

\subsection{Forecast evaluation}
\label{sec:evaluation}

For the historical evaluation, we keep countries and territories with reported TFR estimates covering the 15-year forecast period and predictions from all five models at each of the three origins. This results in a common sample of 144 populations, with all models evaluated against the same observations at each origin. Keeping this sample fixed allows us to compare performance across models and forecast periods without changes in the populations represented.

We assess point forecast accuracy using the median forecast from each model. Because fertility levels differ across populations, we use the symmetric mean absolute percentage error (sMAPE) as our main measure, expressing absolute errors as percentages of the average of the forecasted and observed TFR. We also report mean absolute error (MAE) to show the size of forecast errors in births per woman. We calculate both measures over the fifteen forecast years for each population and origin.

We evaluate the 80\% and 95\% prediction intervals using the Winkler score, empirical coverage, and mean interval width, examining how these measures vary across forecast horizons and fertility groups. The Winkler score adds a penalty to the interval width when an observation falls outside the predicted bounds. The farther the observation lies from the nearest bound, the larger the penalty. Lower scores thus reflect a better balance between width and coverage. Empirical coverage is the proportion of observations within the interval. We also calculate mean quantile loss across all five predicted quantiles on the original TFR scale. To assess whether forecasts tend to overestimate or underestimate fertility, we also calculate the mean difference between the median forecast and observed TFR, and the proportions of observations falling below and above the prediction intervals.

For the overall comparison, we average each population's scores over the fifteen forecast years and the three origins, then report the median across populations. For overall interval coverage and width, we report means across populations, with equal weight for each population.

We compare models by calculating the difference between their scores within each population. We summarize these paired differences with the Hodges--Lehmann estimate, the median of their pairwise averages. For each metric, we use a two-sided Wilcoxon signed-rank test and apply Holm's adjustment across the four comparisons between NeuralTFR and its benchmarks. We compare the spline with BayesTFR in the same way, using a single test for each metric. The estimates describe how large the differences are, while the tests indicate the statistical evidence for them. To calculate confidence intervals, we first average scores across origins and then resample the 144 population-level paired differences with replacement 2,000 times. We use the 2.5th and 97.5th percentiles of the estimates from these resamples as the limits of the 95\% intervals, without adjusting for multiple comparisons. This calculation assumes independent populations and keeps the fitted forecasts fixed. It captures variation across populations, but excludes the uncertainty that would arise from refitting the models. We repeat the estimation separately at each origin to see whether the differences persist across evaluation periods. For comparisons within the fertility groups defined in Section~\ref{sec:design}, we average each population's scores over the origins at which it belonged to the group.

\subsection{Forward projections}
\label{sec:forward_projections}

For the forward projections, we fit NeuralTFR and the spline model using the same data through 2023 and project fifteen years ahead, to 2038. We compare both models with the United Nations World Population Prospects 2024 \citep{un2024wpp} and the GBD 2021 fertility projections, published in 2024 \citep{gbd2024global}. We include the 130 of our 230 populations that have a reported TFR in 2023, the final year of the input history, and are covered by both published systems. The comparison follows all three systems over 2024--2038, although their projections start in different years: 2024 for the United Nations and 2022 for GBD. We retain the published systems' own baseline estimates. Differences between projections consequently include differences in their starting values as well as in the fertility change they anticipate.

For 2038, we report a mean of country TFRs using the United Nations' 2023 population estimates as fixed weights. A change in this mean then reflects fertility change, since the population composition used for weighting stays constant. We count how many central projections fall below 1.5 births per woman to describe the spread of very low fertility, and how many fall below 1.3 to describe its extension to still lower levels. We also group populations by their reported TFR in 2023 and calculate the mean projected change from these values within each group, without population weights. Projections for individual populations can be explored in an interactive visualizer (\url{https://famori.github.io/NeuralTFR/}).

\section{Results}
\label{sec:results}

\subsection{Overall forecasting performance over fifteen years}
\label{sec:overall_performance}

\begin{figure}[!htbp]
\centering
\includegraphics[width=\linewidth]{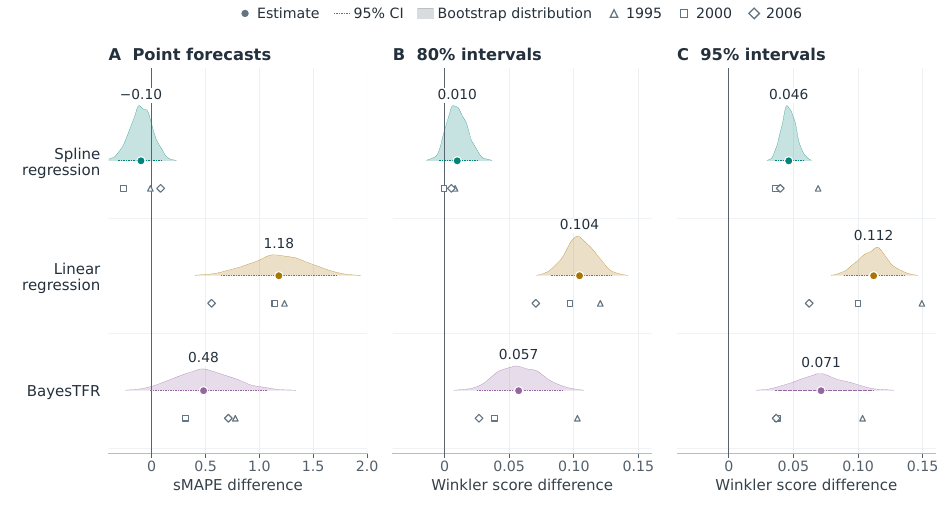}
\caption{\textbf{Paired differences in historical forecasting performance.} Each row compares a benchmark with NeuralTFR for the same 144 populations, with each population's scores averaged over fifteen forecast years and the 1995, 2000, and 2006 origins. Values are the benchmark's score minus NeuralTFR's, so positive values favor NeuralTFR. Dots mark Hodges--Lehmann estimates, lines their 95\% confidence intervals (CI), and shaded curves the distributions of 2,000 bootstrap estimates; open symbols show the estimates for each origin separately. Confidence intervals are unadjusted for multiple comparisons. Panel (A) compares the symmetric mean absolute percentage error (sMAPE), in percentage points; panels (B) and (C) compare Winkler scores for the 80\% and 95\% prediction intervals, in births per woman. Drift, which performs much worse on all three measures, is reported in Table~\ref{tab:paired_comparisons}.}
\label{fig:paired_differences}
\end{figure}

NeuralTFR and the spline have similar point forecast accuracy over the three historical origins. Averaging scores across origins for each population, we obtain median sMAPE values of 8.37\% for NeuralTFR and 8.30\% for the spline. The corresponding medians are 8.96\% for BayesTFR, 9.71\% for the linear regression, and 12.60\% for drift (Figure~\ref{fig:historical_performance}). The paired comparison in Figure~\ref{fig:paired_differences} also shows little difference between NeuralTFR and the spline. Against BayesTFR, NeuralTFR's paired sMAPE advantage is 0.48 percentage points, equivalent to about 5\% of BayesTFR's median score. On the TFR scale, the paired reduction in mean absolute error is 0.010 births per woman; BayesTFR's median error is 0.173. The evidence for a difference in their sMAPE scores is less clear after adjustment for multiple comparisons (Holm-adjusted $p = 0.12$).

For the 80\% intervals, NeuralTFR has lower Winkler scores than BayesTFR, the linear regression, and drift, while the comparison with the spline shows no clear difference. Its 95\% scores are lower than those of all four benchmarks. The advantage over the spline is small: the paired difference amounts to about 4\% of the spline's median Winkler score, although NeuralTFR scores better in 82\% of populations. Against BayesTFR, the paired advantages at the two interval levels amount to 6--7\% of its median Winkler scores (Table~\ref{tab:paired_comparisons}). The absence of a clear difference between NeuralTFR and the spline also extends to mean absolute error and mean quantile loss.

Examining each origin separately (open symbols in Figure~\ref{fig:paired_differences}) shows that neither NeuralTFR nor the spline consistently has better point accuracy. Their 80\% Winkler differences stay close to zero at all three origins, whereas the 95\% scores favor NeuralTFR at every origin. The largest interval advantages over BayesTFR and the linear regression occur in the forecasts from 1995. Thus, the spline recovers much of NeuralTFR's performance in point forecasts and 80\% intervals. The modest neural advantage for 95\% intervals holds across origins. Compared directly with BayesTFR, the spline has better point accuracy and 80\% interval scores, by about 7\% of BayesTFR's median scores, but not clearly better 95\% scores (Table~\ref{tab:spline_bayestfr}).

The overall comparison with BayesTFR conceals a clear difference at very low fertility ($\mathrm{TFR} < 1.5$). In the 40 populations that started at least one forecast period in this group, NeuralTFR's paired sMAPE advantage is 1.76 percentage points, about a fifth of BayesTFR's median score, compared with 5\% in the full sample, and it favors NeuralTFR at all three origins; the spline performs similarly. BayesTFR's forecasts for these populations were too high, by 0.08 births per woman on average and 0.12 in the last five forecast years, while NeuralTFR's errors were close to zero on average. BayesTFR was more accurate mainly where fertility rose strongly after the origin, but as Section~\ref{sec:fertility_change} shows, most of these increases have not lasted. Table~\ref{tab:group_comparisons} reports the paired differences in point accuracy and interval scores, with confidence intervals, for all three fertility groups.

NeuralTFR, the spline, and BayesTFR all have pooled coverage close to the stated interval levels, but the observations outside their intervals are distributed differently between the two tails. In populations starting with moderately low fertility ($1.5 \leq \mathrm{TFR} < 2.1$), 19.6\% of observations lie below BayesTFR's 80\% intervals, almost twice the expected share. Among those with very low fertility ($\mathrm{TFR} < 1.5$), NeuralTFR's 80\% interval misses are balanced, whereas observations outside the linear regression's and drift's intervals fall predominantly above them. At or above replacement, all three leading models underestimate fertility on average: their median forecasts are 0.11 births per woman below observed values for NeuralTFR and the spline, and 0.09 below for BayesTFR. Appendix~\ref{app:interval_groups} provides the detailed interval diagnostics.

\subsection{What the models learn about fertility change}
\label{sec:fertility_change}

\begin{figure}[!htbp]
\centering
\includegraphics[width=\linewidth]{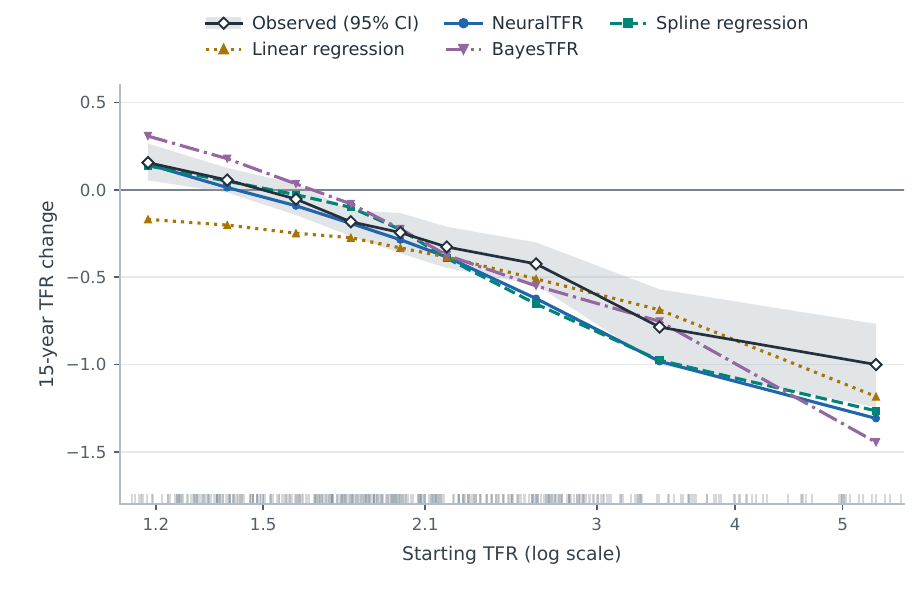}
\caption{\textbf{Observed and forecast fertility change over fifteen years, by fertility at the origin.} Lines connect mean changes across population--origin pairs, using the same 144 populations at the 1995, 2000, and 2006 origins; each pair has equal weight and contributes to every model. Change is measured from reported total fertility rate (TFR) at the origin to either reported TFR fifteen years later or the model's median forecast, in births per woman. Pairs are grouped by starting TFR into ranges: below 1.3, ranges 0.2 wide from 1.3 to 2.3, then 2.3--3, 3--4, and 4 or above. Each range includes its lower bound, contains between 30 and 85 pairs, and is plotted at the mean starting TFR of its pairs, on a logarithmic scale. Linear interpolation between adjacent ranges gives the levels at which change turns negative, reported in the text. The shaded band gives 95\% confidence intervals for observed change from 5,000 bootstrap resamples of populations, keeping all three origins of each sampled population together and treating populations as independent.}
\label{fig:fertility_change}
\end{figure}

Figure~\ref{fig:fertility_change} compares the fertility change the models predicted with the change observed after each historical origin, plotting mean change against fertility at the origin. NeuralTFR and the spline regression predict modest increases, or little change, on average for populations starting below 1.5 births per woman, and progressively larger declines at higher starting levels. BayesTFR also predicts recovery below 1.5, but the average increases are larger. The linear regression predicts decline at every level, including below 1.5, where fertility subsequently rose on average. Observed change turned negative at about 1.5 births per woman, close to where the lines for NeuralTFR (1.42) and the spline (1.53) cross zero, whereas BayesTFR's forecasts turn negative only at 1.66 and imply small increases where fertility declined on average. The similarity between NeuralTFR and the spline therefore extends beyond forecast accuracy to the pattern of fertility change they predict.

The leading models share errors at both ends of the fertility range. At the lowest levels, the pooled means combine increases after 1995 and 2000, mostly in European populations, with a failure at the 2006 origin: fertility fell in the four populations starting below 1.2 births per woman (Hong Kong, Macao, Taiwan, and South Korea), where NeuralTFR, the spline, and BayesTFR all predicted increases (Figures~\ref{fig:country_trajectories} and~\ref{fig:additional_trajectories_decline}). Many of the increases at the lowest levels were reversed after the evaluation period. Of the 60 population--origin pairs starting below 1.5 in which fertility was higher fifteen years later, 51 had fallen below that level by the most recent observation, and 31 below their level at the origin. Several recoveries peaked in 2015 or 2021, as the evaluation periods from 2000 and 2006 ended, so the fifteen-year comparison records them close to their highest point. At or above replacement, all four models forecast larger declines than observed on average.

\subsection{Forecasts for individual populations}
\label{sec:country_trajectories}

\begin{figure}[!htbp]
\centering
\includegraphics[width=\linewidth]{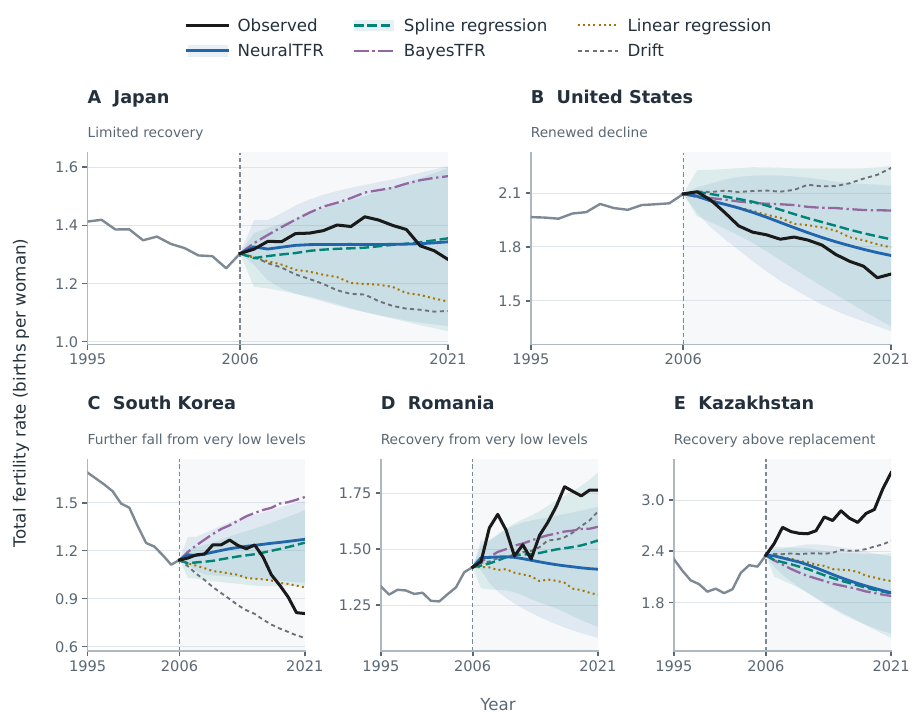}
\caption{\textbf{Individual fertility trajectories and forecasts from 2006.} Five cases selected after examining observations and forecasts to illustrate contrasting patterns within the common 144-population sample. Gray lines show reported history before the origin; black lines show observations from 2006 through 2021. Colored lines show predictive medians from the five models; blue and green shading show the 80\% prediction intervals for NeuralTFR and the spline. Vertical dashed lines mark the forecast origin, with forecast paths joined to the last observation. Vertical scales vary across panels. Appendix~\ref{app:country_trajectories} shows additional cases.}
\label{fig:country_trajectories}
\end{figure}

Figure~\ref{fig:country_trajectories} shows how the patterns described above appear in individual forecasts. It compares forecasts from the 2006 origin with the fertility observed over the following fifteen years in five populations, selected after examining observations and forecasts to illustrate contrasting patterns. These cases show how the models behave, not how well they perform, which Section~\ref{sec:overall_performance} assesses across all 144 populations. Appendix~\ref{app:country_trajectories} presents additional cases.

Japan illustrates how modest recovery can coexist with persistent very low fertility. Fertility rose temporarily after 2006 but was close to its starting level again by 2021. NeuralTFR projects a small increase that levels off, while BayesTFR projects a larger recovery that continues. NeuralTFR does not reproduce the rise and subsequent fall. Its forecast nevertheless stays closer to the observed trajectory over the evaluation period because the projected increase remains small. The spline produces a similarly modest increase, showing that the nonlinear relationship estimated from pooled histories also captures this behavior.

Recovery in the recent past can also precede another decline. In the United States, fertility had risen toward replacement before the forecast origin, then fell through most of the following period. NeuralTFR captures more of the decline than BayesTFR or the spline, although it still underestimates its extent. Both cases illustrate why forecasts can benefit from allowing recovery to remain modest or to give way to a further decline.

Forecasting stabilization or modest recovery at very low fertility can, however, produce errors in either direction. NeuralTFR and the spline anticipate too little decline in South Korea, missing its subsequent fall below one birth per woman, and too little recovery in Romania. BayesTFR follows Romania's recovery more closely.

At higher starting levels, forecasts of decline can fail when fertility remains high or rises. All models except drift predict decline in Kazakhstan, despite the recovery already underway at the forecast origin. Drift captures the direction of change better in this case, although it underestimates the increase.

\subsection{Projected fertility to 2038}
\label{sec:forward_results}

\begin{figure}[!htbp]
\centering
\includegraphics[width=\linewidth]{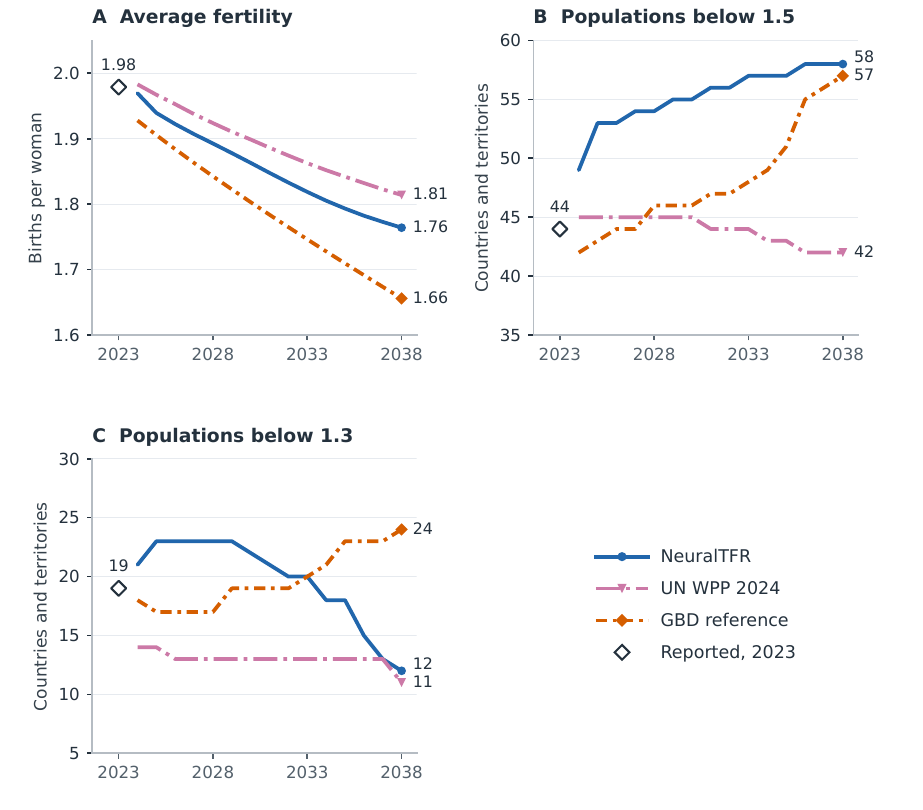}
\caption{\textbf{Projected fertility to 2038.} NeuralTFR, fitted to observations through 2023 and projecting fifteen years ahead, compared with the median projections of the United Nations World Population Prospects 2024 (UN WPP 2024), which are based on BayesTFR, and the reference scenario of the Global Burden of Disease (GBD) 2021 study, for the 130 countries and territories covered by all three systems. Panel (A) shows the mean of country total fertility rates (TFR) weighted by fixed 2023 population. Panels (B) and (C) count the populations whose central projection lies below 1.5 and 1.3 births per woman. Open diamonds show reported values in our empirical dataset in 2023; the published systems retain their own baseline estimates. Labels at the right give values in 2038.}
\label{fig:forward_projections}
\end{figure}

Across the 130 countries and territories included in the comparison, all three systems project declining population-weighted mean fertility (Figure~\ref{fig:forward_projections}). NeuralTFR projects a decline from the reported mean of 1.98 births per woman in 2023 to 1.76 in 2038, between the United Nations' 1.81 and GBD's 1.66. In 2024, the first year common to the three projection series, their means are 1.97, 1.98, and 1.93, respectively. Subsequent declines to 2038 are therefore 0.21 births per woman for NeuralTFR, 0.17 for the United Nations, and 0.27 for GBD. The differences in 2038 reflect both differences already present in 2024 and the changes projected thereafter.

NeuralTFR projects very low fertility in about as many populations as GBD, but substantially fewer populations below 1.3 births per woman. In 2023, 44 of these populations had a TFR below 1.5. By 2038, NeuralTFR projects 58 below this level, close to the 57 projected by GBD and well above the 42 projected by the United Nations. NeuralTFR already projects 57 by 2033, whereas most of GBD's increase comes after that year (Figure~\ref{fig:forward_projections}B). Below 1.3, NeuralTFR projects 12 populations, fewer than the 19 in 2023 and similar to the United Nations' 11, whereas GBD projects 24.

The spline regression, which achieved similar historical point accuracy to NeuralTFR, projects a similar pattern, with a weighted mean of 1.75 and 10 populations below 1.3. It projects fewer populations below 1.5, 50 compared with NeuralTFR's 58, but this difference reflects projections close to the threshold. Nine populations fall below 1.5 only in NeuralTFR's central projections, and one only in the spline's; in these ten populations, both models project values between 1.43 and 1.53. The broader projected pattern is therefore not specific to the recurrent architecture.

Among the 44 populations starting below 1.5 births per woman, NeuralTFR projects a modest increase in fertility on average, but all remain below this level in its central projections for 2038, compared with 35 in the United Nations projections. Of the 43 populations with moderately low fertility in our 2023 data, NeuralTFR projects fourteen to fall below 1.5 by 2038 and the United Nations seven. Their mean projected fertility levels in 2038 are respectively 0.15 and 0.05 births per woman below the reported 2023 values in our dataset.

\section{Discussion}
\label{sec:discussion}

The question motivating this paper is whether recent fertility declines will give way to stabilization and recovery, and how widely very low fertility will spread. Read through models that learn the form of fertility change from pooled country histories, the historical record points to an answer that combines both possibilities. In the forward comparison, our two models, NeuralTFR and the spline regression, combine modest recovery from the lowest levels with continued decline from moderately low levels, so that very low fertility becomes more widespread by 2038 without becoming much deeper. Both project very low fertility in more populations than the United Nations, NeuralTFR in about as many as GBD. They also project few populations below 1.3 births per woman, similar to the United Nations and far fewer than GBD. The main difference from the United Nations is therefore how widely very low fertility spreads; the main difference from GBD is how far it falls.

The difference from the United Nations can be traced to the fertility level at which projected recovery gives way to decline. In the historical evaluation, observed fifteen-year change turned negative at about 1.5 births per woman. NeuralTFR and the spline place this transition close to that level. BayesTFR places it at about 1.66 in our historical fits, implying stabilization or recovery for populations between the two levels. In this range, BayesTFR's historical forecasts were too high: observations fell below its 80\% intervals about twice as often as expected. The forward projections also differ most here, with NeuralTFR projecting more populations moving from moderately low to very low fertility, and larger average declines among them, than the United Nations.

Historical experience supports some parts of this pattern more than others. Many populations have passed through moderately low fertility, and the models' forecasts in this range were close to observed change on average, so continued decline from these levels follows the best-supported part of the learned relationship. At higher fertility levels, the leading models forecast faster declines than those observed, consistent with the difficulty of anticipating stalls in fertility transitions from past trajectories alone \citep{bongaarts2006causes}.

The limited depth of projected decline depends instead on modest recovery at the lowest levels, where historical observations are few. The increases observed at these levels come mainly from European populations whose fertility rose after reaching its lowest levels around 2000, partly because postponement slowed, a source of increase that need not recur elsewhere \citep{goldstein2009end,bongaarts2012demographic}. As noted in the introduction, most of these increases have since been reversed, including in 11 of the 15 populations in which they made BayesTFR's historical forecasts more accurate than NeuralTFR's. Populations that have reached very low fertility since then, in East Asia and elsewhere, have on average continued to decline, but too few years have passed for these experiences to shape the long-horizon forecasts. Projected recovery at the lowest levels, which keeps few populations below 1.3, is therefore the least certain part of the projected pattern, and the reversals suggest that it is more likely to prove too high than too low.

Methodologically, the results show what learning the form of fertility change adds, and how much of it can be expressed more transparently. Across fifteen-year forecasts from three historical origins, NeuralTFR is competitive with BayesTFR: its sMAPE is similar overall but lower where fertility is very low, and its Winkler scores for 80\% and 95\% intervals are slightly lower. Still, interpreting these forecasts did not reveal complex dependencies between the detailed shape of past trajectories and future change. What emerged was a simple relationship: fertility tends to rise modestly from the lowest levels and keep declining from higher ones. The recent trajectory affects the size of these changes more than their direction. Once this relationship is represented in a pooled spline regression, the regression matches NeuralTFR's point accuracy and 80\% interval performance and projects a similar pattern to 2038. In this setting, much of the forecasting value lies in pooling country histories and allowing fertility change to depend nonlinearly on its level. This is consistent with evidence that models estimated jointly across many series can be simple yet competitive \citep{montero2021principles}, and that forecast accuracy for completed cohort fertility hardly improves with method complexity \citep{bohkewald2018forecast}. It is also closer to recent evidence from mortality forecasting \citep{park2025next} than to settings in which recurrent architectures add accuracy beyond pooling \citep{hewamalage2021recurrent}.

What the regression does not fully recover is also informative. NeuralTFR's 95\% intervals keep slightly lower Winkler scores than the spline's at all three origins. For populations with very low fertility, the spline matched NeuralTFR's long-horizon 80\% intervals only after we smoothed its coefficients across forecast years. The improvement also depends on the strength of the penalty. When estimated separately for each year, quantiles at distant horizons rely on fewer training windows, especially at low fertility levels. With smoothing, estimates for distant years can draw on the information available at nearer horizons. The recurrent decoder already does this sharing across years through a common set of parameters for all its quantiles. That feature may help explain its remaining advantage.

Two limitations of the evaluation are relevant here. First, the historical origins provide less experience of very low fertility than is available when fitting the forward projections. Second, the forecast periods overlap, and populations share regional and global influences, so a model that misses a common change, such as a regional fertility decline, misses it in several populations at once. Our bootstrap intervals and significance tests treat populations as independent and may therefore overstate the precision of the comparisons. We therefore judge differences by their size and by whether they point in the same direction at all three origins, rather than by statistical significance alone.

Our projections extend the relationships observed in country histories over the next fifteen years, conditional on those relationships continuing. These relationships help forecast fertility, but they do not explain what produces it. Fertility histories alone cannot tell us why populations at similar levels follow different paths, for instance toward sustained recovery or renewed decline. Nor can they show how changes in social and institutional conditions would alter these paths, or anticipate changes without historical precedent. In Hong Kong, Macao, Taiwan, and South Korea, all three leading models expected increases after 2006, but fertility kept falling as marriage was delayed or forgone. Unlike in much of Europe, later marriage there was not offset by cohabitation or childbearing outside marriage \citep{jones2007delayed}. Combining fertility histories with selected covariates, such as measures of birth timing, education, or family policy, or with regional groupings that let populations within a region share patterns of change, is a natural next step. More broadly, we believe that models that learn from the historical record should be seen not as a replacement for established projection systems, but as a complement to them: by making explicit what past trajectories suggest about future change, they provide a baseline against which both the assumptions built into these systems and the value of additional information can be assessed.

\section{Conclusions}
\label{sec:conclusions}

The disagreement among projection systems is often framed as a choice between recovery and continued decline. Read through the dynamics learned from it, the historical record suggests that both occur, but in different places: modest recovery where fertility is already lowest does not prevent very low fertility from spreading to populations now just above it. A recurrent neural network and a pooled spline regression recover this pattern from the same country histories. More broadly, our results show how learning the form of fertility change from the historical record, rather than assuming it, can bring empirical evidence to a debate that has so far rested largely on assumptions about fertility after the transition.

\section*{Data and code availability}

The repository accompanying this article (\url{https://github.com/FaMori/NeuralTFR}) contains the fertility panel with the source of each estimate, the external inputs from which it was built, and also the code that fits NeuralTFR, the drift and regression benchmarks at every forecast origin and re-estimates BayesTFR at historical origins. It includes as well the trained networks and the recorded forecasts of all models, so that each result and figure reported in the article can be reproduced. Projections for individual populations can be explored in the interactive visualizer at \url{https://famori.github.io/NeuralTFR/}.

\clearpage
\appendix
\numberwithin{figure}{section}
\numberwithin{table}{section}
\numberwithin{equation}{section}
\section{Fertility composition of the panel}
\label{app:fertility_composition}

\begin{figure}[!htbp]
\centering
\includegraphics[width=\textwidth]{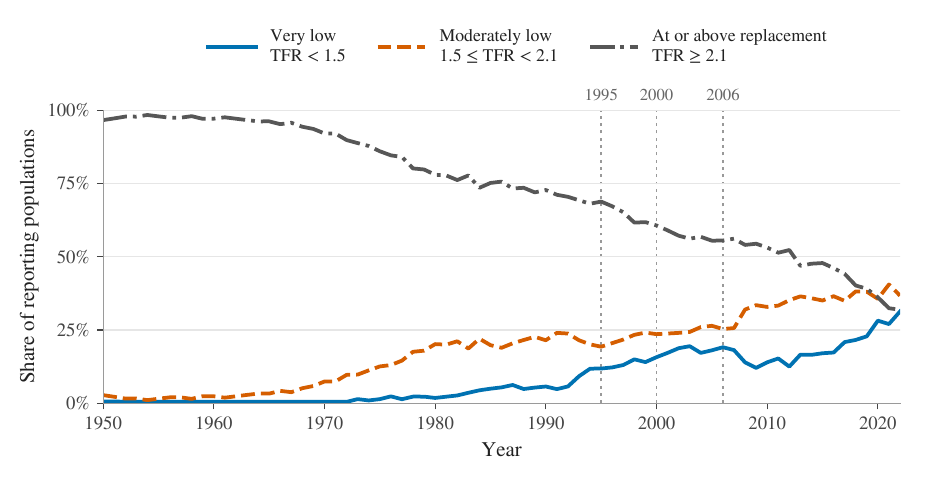}
\caption{\textbf{Fertility composition of the panel over time.} Shares of reporting countries and territories in different total fertility rate (TFR) groups. Interpolated values excluded. Shares shown only for years with at least 150 reporting populations (1950--2022). Vertical dotted lines mark the historical forecast origins: 1995, 2000, and 2006.}
\label{fig:coverage_composition}
\end{figure}

\clearpage
\section{Pooled quantile-regression models}
\label{app:pooled_regressions}
The transformation of fertility in Section~\ref{sec:neuraltfr} uses $a=1.3$ and $b=0.55$, rounded values close to the mean and standard deviation of log TFR in the panel. Let $\Delta z_{i,t}=z_{i,t}-z_{i,t-1}$ denote the annual change in transformed fertility. Both regressions use the current level and the $k$ most recent annual changes, $\boldsymbol{v}_{i,t}=(z_{i,t},\Delta z_{i,t},\ldots,\Delta z_{i,t-k+1})$. For each forecast year $h$, we estimate five functions, one for the median and four for the distances between successive quantiles,
\begin{equation}
\eta_{h,r}(\boldsymbol{v})=\alpha_{h,r}+\boldsymbol{\beta}_{h,r}^{\mathsf T}\boldsymbol{\phi}(\boldsymbol{v}),
\qquad r=0,\ldots,4,
\label{eq:pooled_predictors}
\end{equation}
with coefficients shared across countries. In the linear regression, $\boldsymbol{\phi}$ contains the predictors themselves. In the spline regression, each predictor enters as a linear spline with knots at its nine training-window deciles, so that its slope can change across its range; the predictors contribute additively, without interactions. For each predictor $v_j$, the unstandardized spline terms are
\[
\left[v_j,\;(v_j-\kappa_{j,1})_+,\;\ldots,\;(v_j-\kappa_{j,9})_+\right],
\]
where $(u)_+=\max(u,0)$ and the knots $\kappa_{j,1},\ldots,\kappa_{j,9}$ are its nine training-window deciles. Each term is then centered and scaled using its training mean and standard deviation. We model the median directly and pass the other four functions through a softplus transformation, which keeps the distances between successive quantiles positive, so that the quantiles cannot cross.

We estimate both regressions from the pooled training windows by minimizing mean quantile loss on the transformed scale (Section~\ref{sec:neuraltfr}). For the spline, we add a penalty on the second differences of the coefficients across adjacent forecast years,
\begin{equation}
\mathcal{L}_{\mathrm{spline}}=\mathcal{L}_{\mathrm{quantile}}+
\lambda\sum_{h=2}^{14}\sum_{r=0}^{4}
\left\|\boldsymbol{\theta}_{h+1,r}-2\boldsymbol{\theta}_{h,r}+\boldsymbol{\theta}_{h-1,r}\right\|_2^2,
\label{eq:pooled_smoothing}
\end{equation}
where $\boldsymbol{\theta}_{h,r}$ collects the intercept and coefficients of function $r$ in year $h$. In other words, a coefficient may stay constant or change at a constant rate across forecast years without penalty, but changes in that rate are penalized. Both models are fitted with 4,000 full-sample Adam updates, starting from zero coefficients, with a learning rate of 0.05 reduced to zero on a cosine schedule.

We chose the number of recent changes and the penalty using the ten-year validation periods before each historical origin, with training data ending in 1985, 1990, and 1996. We considered $k\in\{0,1,3\}$ for the linear regression and $k\in\{0,1,2,3,11\}$ for the unpenalized spline; for both, validation loss on the TFR scale, summed across origins, is lowest at $k=3$. With $k=3$, we then compared $\lambda\in\{0,10^{-4},10^{-3},10^{-2},10^{-1},1,10\}$. Validation loss selects $\lambda=0.1$ at each origin separately, when summed across origins, and when any one origin is left out, so the choice does not rely on validation periods that overlap another origin's test years. We refit both models at each origin with these settings, recalculating the knots and scaling from its training windows, and keep the same settings for the projections from 2023.

We developed the regressions after examining NeuralTFR's responses and its historical forecasts. Although validation determines the penalty, the comparison is therefore not fully independent of results we had already seen. Validation also includes only populations with the later observations needed for complete test forecasts; these observations decide which populations are included but are never used as predictors or targets.

\subsection{Sensitivity to the smoothing penalty}
\label{app:spline_sensitivity}

Validation losses summed across origins differ little between penalties: 0.16623 with $\lambda=0.1$, 0.16641 with $\lambda=0.01$, and 0.16720 without smoothing. When we resample populations 5,000 times, keeping each population's observations together across origins, $\lambda=0.1$ is preferred in 71.8\% of resamples and $\lambda=0.01$ in 23.2\%. The choice matters most for long-horizon intervals at very low fertility. For populations starting below 1.5 births per woman, the 80\% intervals cover 79.6\% of observations in years 11--15 with $\lambda=0.1$, against 64.7\% with $\lambda=0.01$, 61.8\% without smoothing, and 80.0\% for NeuralTFR. Smoothing therefore improves long-horizon coverage, but by an amount that depends on the penalty. The forward results hardly change: with either $\lambda=0.1$ or $\lambda=0.01$, the spline projects 50 of the 130 populations below 1.5 and ten below 1.3 in 2038, with a population-weighted mean of 1.75.

\clearpage
\section{Post hoc interpretability analyses}
\label{app:interpretability}
We use two exercises to see how NeuralTFR's forecasts depend on current fertility and on the path that led to it. Both pass modified histories through the fitted networks without changing their parameters.

The first uses simulated twelve-year histories that end at a fertility level $\ell$ after changing by a constant $s$ births per woman each year,
\begin{equation}
\mathrm{TFR}_{t-j}=\ell-sj,
\qquad j=0,\ldots,11.
\label{eq:interpretability_history}
\end{equation}
Combining seven levels with four annual changes (Table~\ref{tab:interpretability_grid}) gives 28 histories, which we transform as in forecasting, annual changes included. For each history, we take the median forecast across the ten fits at each origin and express it as a change from $\ell$. Histories that end at the same level but arrive there differently show the effect of the preceding trend, while histories with the same trend show how the forecast changes with the level.

\begingroup
\setlength{\intextsep}{6pt}
\begin{table}[!htbp]
\centering
\caption{Values used in the controlled-input grid.}
\label{tab:interpretability_grid}
\begin{tabular}{@{}ll@{}}
\hline
Dimension & Values \\
\hline
Fertility level, $\ell$ & $0.7,\ 0.9,\ 1.2,\ 1.6,\ 2.1,\ 3.0,\ 4.5$ \\
Annual change, $s$ & $-0.08,\ -0.04,\ 0,\ 0.04$ \\
Forecast horizon, $h$ & $1,\ 5,\ 10,\ 15$ years \\
\hline
\end{tabular}
\end{table}
\endgroup

At each origin, the forecasts imply increasing fertility over fifteen years for histories ending at 1.2 births per woman or below and declining fertility for those ending at 1.6 or above, so the change of direction lies between these two levels of the grid. The preceding trend alters the size of the forecast change, mainly at the highest level, but never its direction. Below 1.0 births per woman, the forecasts are largely extrapolations, since only two populations had reached such levels before 2006. Because these histories have no observed future, the results describe how the fitted network behaves over this range and horizon, not a general threshold for recovery.

The second exercise asks how much the preceding trajectory adds to forecast accuracy. For the 144 populations at the three historical origins, we replace each observed history with a constant sequence at its final value, so that all annual changes are zero, and compare the resulting forecasts with those from the observed histories against the same outcomes, using mean quantile loss on the TFR scale. Removing the trajectory increases loss by 2.3\% overall, by 4.2\%, 1.6\%, and 1.1\% at the 1995, 2000, and 2006 origins, and in 29 of the 30 individual fits. The increase is about 10\% in the first two forecast years and close to zero by year fifteen, so the trajectory matters mainly for near-term forecasts.

\clearpage
\section{Prediction interval performance}
\label{app:interval_groups}
On average across populations, the 80\% and 95\% intervals of NeuralTFR, the spline, and BayesTFR contain close to the stated share of observations. Coverage nevertheless varies widely from one population to another: NeuralTFR's 95\% intervals, for instance, contain every observation in 57\% of populations, but fewer than 80\% of observations in 6\%.

Where the intervals miss, they miss in different directions. Observed fertility falls above the intervals of NeuralTFR and the spline more often than below them, whereas it falls below BayesTFR's more often than above. For BayesTFR, these misses are concentrated among populations starting with moderately low fertility, where its forecasts were too high (Section~\ref{sec:overall_performance}).

\begin{figure}[!htbp]
\centering
\includegraphics[width=\linewidth]{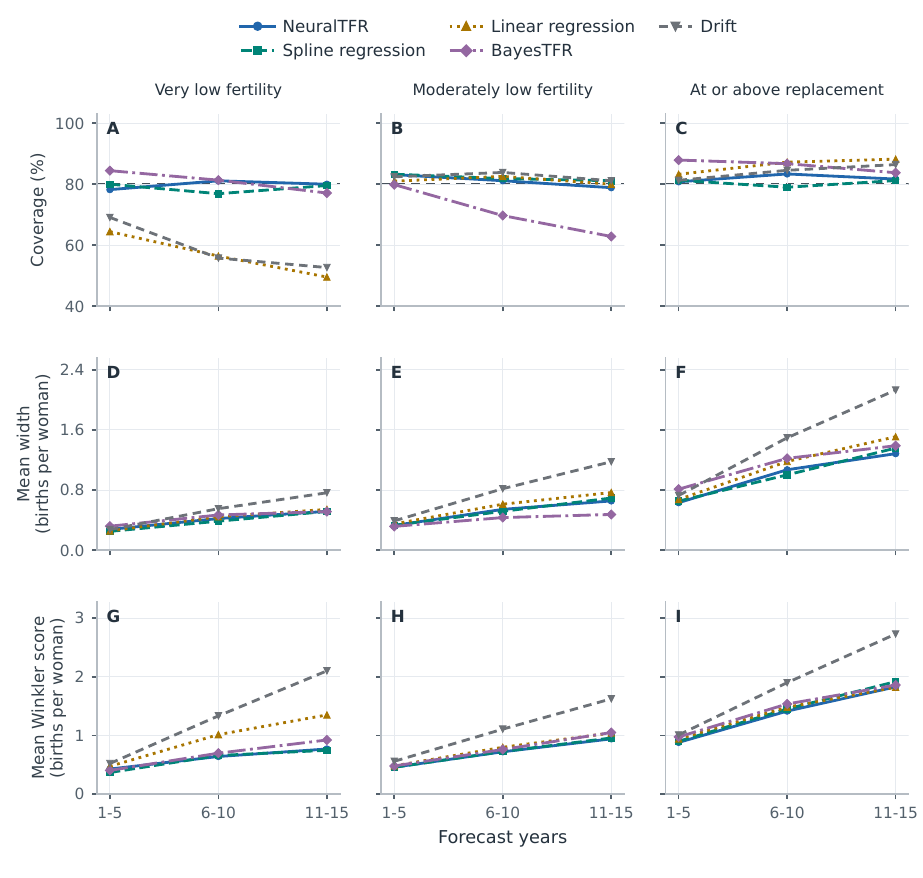}
\caption{\textbf{Performance of 80\% prediction intervals by starting fertility and forecast horizon.} Columns show the three groups defined by total fertility rate (TFR) at the origin: very low ($\mathrm{TFR}<1.5$), moderately low ($1.5\leq\mathrm{TFR}<2.1$), and at or above replacement ($\mathrm{TFR}\geq2.1$), with 90, 140, and 202 population--origin pairs. Rows show coverage, mean width, and mean Winkler score for years 1--5, 6--10, and 11--15, averaging the same population--origin pairs with equal weights. Dashed lines mark 80\% coverage; lower Winkler scores indicate better intervals.}
\label{fig:interval_groups_80}
\end{figure}

Figure~\ref{fig:interval_groups_80} describes how the 80\% intervals perform in the three fertility groups as the forecast horizon lengthens. At very low fertility, NeuralTFR and the spline keep coverage close to 80\% even in the last five years, with intervals of similar width. At the 95\% level, the spline has the lower mean score in this group, although NeuralTFR scores better in 80\% of these populations: the mean reflects a few large misses by NeuralTFR in Hong Kong and Macao after 1995 and 2000. Appendix~\ref{app:spline_sensitivity} shows the role of smoothing in the spline's long-horizon coverage. At moderately low fertility, BayesTFR's coverage declines with the horizon, to about 63\% in the last five years: its intervals are narrower than those of the other two models, but score worse. At or above replacement, all models come close to the stated coverage at long horizons.

\clearpage
\section{Historical forecasting scores and paired comparisons}
\label{app:historical_performance}

\begin{figure}[!htbp]
\centering
\includegraphics[width=\linewidth]{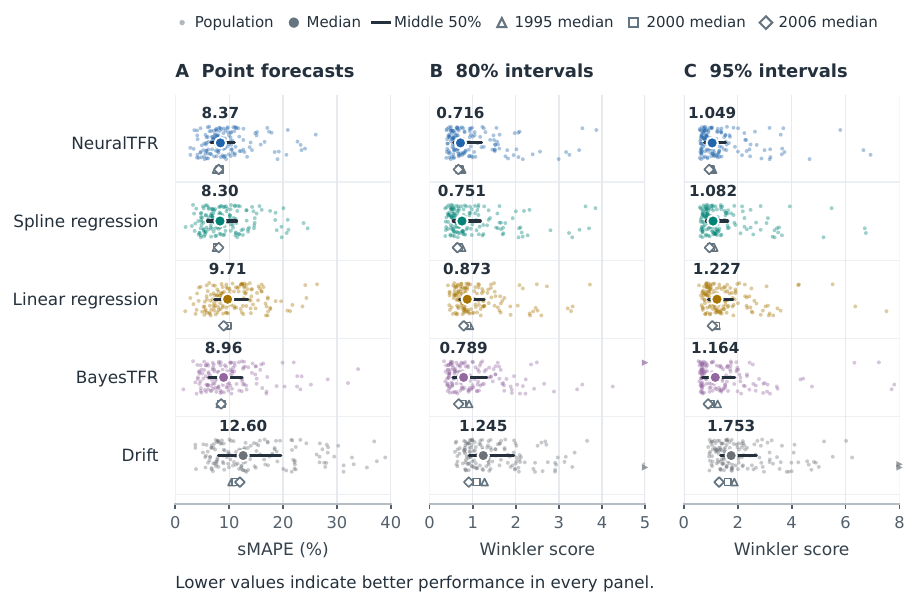}
\caption{\textbf{Historical forecasting performance over fifteen years.} All five models forecast the total fertility rate (TFR) for the same 144 populations at the 1995, 2000, and 2006 origins. Small dots show individual populations, with each population's scores averaged across origins. Large labeled dots give the median across populations; black segments cover the middle 50\%. Open symbols give the median for each origin separately. In each of panels (B) and (C), two populations fall outside the plotted range and are marked by arrowheads at the axis edge. Panel (A) reports the symmetric mean absolute percentage error (sMAPE); panels (B) and (C) report Winkler scores for the 80\% and 95\% prediction intervals, respectively, in births per woman. Lower scores indicate better performance.}
\label{fig:historical_performance}
\end{figure}

\begin{table}[!htbp]
\centering
\caption{\textbf{Paired comparisons with NeuralTFR over fifteen years.} Hodges--Lehmann estimates of the benchmark's score minus NeuralTFR's for the same 144 populations, with each population's scores averaged over the three origins; positive values favor NeuralTFR. Confidence intervals (CI) come from 2,000 bootstrap resamples of populations. ``NeuralTFR better'' gives the percentage of populations in which NeuralTFR has the lower score. Adjusted $p$-values are from two-sided Wilcoxon signed-rank tests, Holm-adjusted across the four benchmarks for each measure (Section~\ref{sec:evaluation}).}
\label{tab:paired_comparisons}
\small
\begin{tabular}{lrcrr}
\toprule
Benchmark & Difference & 95\% CI & NeuralTFR better (\%) & Adjusted $p$ \\
\midrule
\multicolumn{5}{l}{\textit{sMAPE (percentage points)}} \\
\quad Spline regression & $-$0.10 & [$-$0.31, 0.10] & 47 & 0.337 \\
\quad Linear regression & 1.18 & [0.64, 1.71] & 67 & $<$0.001 \\
\quad BayesTFR & 0.48 & [$-$0.01, 1.07] & 52 & 0.120 \\
\quad Drift & 4.79 & [3.70, 5.96] & 77 & $<$0.001 \\
\addlinespace
\multicolumn{5}{l}{\textit{Winkler score, 80\% intervals (births per woman)}} \\
\quad Spline regression & 0.010 & [$-$0.004, 0.026] & 50 & 0.195 \\
\quad Linear regression & 0.104 & [0.083, 0.129] & 82 & $<$0.001 \\
\quad BayesTFR & 0.057 & [0.025, 0.092] & 62 & 0.002 \\
\quad Drift & 0.432 & [0.369, 0.516] & 92 & $<$0.001 \\
\addlinespace
\multicolumn{5}{l}{\textit{Winkler score, 95\% intervals (births per woman)}} \\
\quad Spline regression & 0.046 & [0.036, 0.058] & 82 & $<$0.001 \\
\quad Linear regression & 0.112 & [0.089, 0.136] & 81 & $<$0.001 \\
\quad BayesTFR & 0.071 & [0.036, 0.112] & 67 & $<$0.001 \\
\quad Drift & 0.668 & [0.602, 0.781] & 95 & $<$0.001 \\
\bottomrule
\end{tabular}

\end{table}

\begin{table}[!htbp]
\centering
\caption{\textbf{Paired comparison of the spline regression with BayesTFR over fifteen years.} Differences are BayesTFR's score minus the spline's, so positive values favor the spline; estimates and intervals are calculated as in Table~\ref{tab:paired_comparisons}, in percentage points for sMAPE and births per woman for Winkler scores. ``Spline better'' gives the percentage of populations in which the spline has the lower score. The $p$-values are from two-sided Wilcoxon signed-rank tests, one for each measure.}
\label{tab:spline_bayestfr}
\small
\begin{tabular}{lrcrr}
\toprule
Measure & Difference & 95\% CI & Spline better (\%) & $p$ \\
\midrule
sMAPE & 0.60 & [0.14, 1.06] & 58 & 0.012 \\
Winkler score, 80\% intervals & 0.055 & [0.018, 0.092] & 58 & 0.005 \\
Winkler score, 95\% intervals & 0.022 & [$-$0.020, 0.067] & 49 & 0.320 \\
\bottomrule
\end{tabular}

\end{table}

\clearpage
NeuralTFR's point-forecast advantage over BayesTFR is largest at very low fertility ($\mathrm{TFR} < 1.5$). Its advantages in 80\% and 95\% Winkler scores are most clearly supported at or above replacement ($\mathrm{TFR} \geq 2.1$), while interval-score differences within the two low-fertility groups are less conclusive.

\begin{table}[!htbp]
\centering
\caption{\textbf{Paired comparison of NeuralTFR with BayesTFR by starting fertility.} Differences are BayesTFR's score minus NeuralTFR's, so positive values favor NeuralTFR. Scores are averaged over fifteen forecast years and then over the origins at which each population belonged to the group. Entries give Hodges--Lehmann estimates and 95\% confidence intervals (CI) from 2,000 bootstrap resamples of populations, retaining each population's origins together; intervals are unadjusted for multiple comparisons. Units are percentage points for sMAPE and births per woman for Winkler scores. Groups are defined by reported TFR at each origin, so a population can contribute to more than one group.}
\label{tab:group_comparisons}
\small
\begin{tabular}{lrc}
\toprule
Measure & Difference & 95\% CI \\
\midrule
\multicolumn{3}{l}{\textit{Very low fertility ($\mathrm{TFR}<1.5$)}} \\
\multicolumn{3}{l}{40 populations; 90 population--origin pairs} \\
\quad sMAPE & 1.76 & [0.26, 3.48] \\
\quad Winkler score, 80\% intervals & 0.029 & [$-$0.020, 0.100] \\
\quad Winkler score, 95\% intervals & 0.022 & [$-$0.010, 0.050] \\
\addlinespace
\multicolumn{3}{l}{\textit{Moderately low fertility ($1.5\leq\mathrm{TFR}<2.1$)}} \\
\multicolumn{3}{l}{69 populations; 140 population--origin pairs} \\
\quad sMAPE & 0.55 & [$-$0.16, 1.27] \\
\quad Winkler score, 80\% intervals & 0.058 & [$-$0.015, 0.139] \\
\quad Winkler score, 95\% intervals & $-$0.033 & [$-$0.076, 0.052] \\
\addlinespace
\multicolumn{3}{l}{\textit{At or above replacement ($\mathrm{TFR}\geq2.1$)}} \\
\multicolumn{3}{l}{79 populations; 202 population--origin pairs} \\
\quad sMAPE & 0.51 & [$-$0.04, 1.25] \\
\quad Winkler score, 80\% intervals & 0.096 & [0.050, 0.139] \\
\quad Winkler score, 95\% intervals & 0.160 & [0.109, 0.201] \\
\bottomrule
\end{tabular}

\end{table}

\clearpage
\section{Additional country trajectories}
\label{app:country_trajectories}
\begin{figure}[!htbp]
\centering
\includegraphics[width=\linewidth]{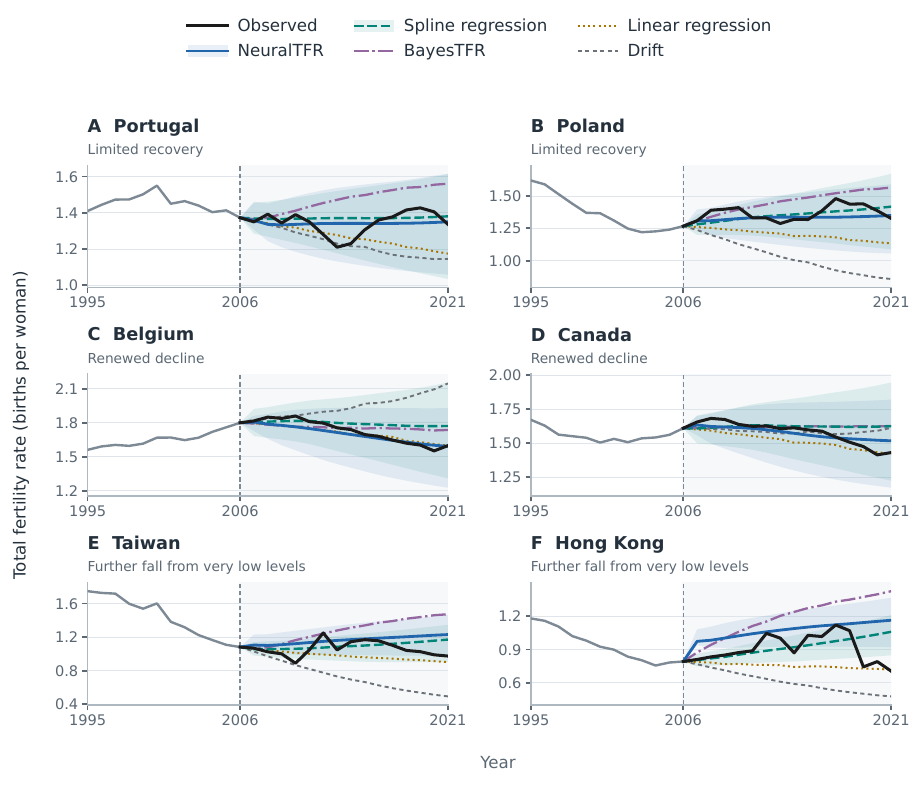}
\caption{\textbf{Additional cases of limited recovery, renewed decline, and further falls at very low fertility.} Lines, intervals, and the forecast boundary follow Figure~\ref{fig:country_trajectories}; vertical scales vary across panels.}
\label{fig:additional_trajectories_decline}
\end{figure}

\begin{figure}[!htbp]
\centering
\includegraphics[width=\linewidth]{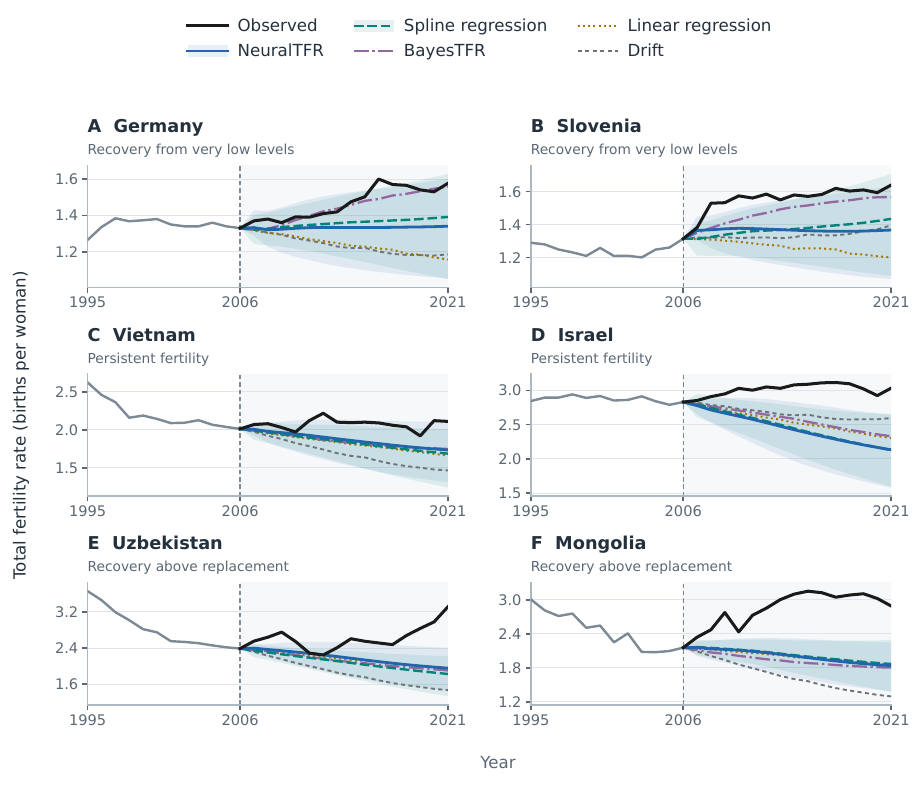}
\caption{\textbf{Additional cases of recovery and of persistent or rising fertility.} Conventions as in Figure~\ref{fig:additional_trajectories_decline}.}
\label{fig:additional_trajectories_recovery}
\end{figure}

Figures~\ref{fig:additional_trajectories_decline} and~\ref{fig:additional_trajectories_recovery} add twelve cases to the five patterns in Figure~\ref{fig:country_trajectories}, using the same 2006 origin and forecast period. As in Figure~\ref{fig:country_trajectories}, we selected them after examining observations and forecasts, so they show that the patterns recur but not how common they are.

Figure~\ref{fig:additional_trajectories_decline} extends the first three patterns. Portugal and Poland resemble Japan: fertility recovers only slightly from very low levels, close to the modest increases projected by NeuralTFR and the spline, while BayesTFR projects a larger recovery that continues. Belgium and Canada resemble the United States, where fertility declines again after rising before 2006. Taiwan and Hong Kong resemble South Korea: fertility ends below its 2006 level, although NeuralTFR, the spline, and BayesTFR all predict recovery.

Figure~\ref{fig:additional_trajectories_recovery} extends the other two. Germany and Slovenia resemble Romania, where fertility recovers from very low levels and BayesTFR captures more of the increase. The remaining cases resemble Kazakhstan. Fertility stays near replacement in Vietnam and near three births per woman in Israel while all five central forecasts imply decline, and it rises in Uzbekistan and Mongolia despite the declines forecast by NeuralTFR, the spline, and BayesTFR.

\clearpage

\clearpage
\bibliographystyle{chicago}
\bibliography{local_biblio}

\begin{thebibliography}{}

\bibitem[\protect\citeauthoryear{Alkema, Raftery, Gerland, Clark, Pelletier,
  Buettner, and Heilig}{Alkema et~al.}{2011}]{alkema2011probabilistic}
Alkema, L., A.~E. Raftery, P.~Gerland, S.~J. Clark, F.~Pelletier, T.~Buettner,
  and G.~K. Heilig (2011).
\newblock Probabilistic projections of the total fertility rate for all
  countries.
\newblock {\em Demography\/}~{\em 48\/}(3), 815--839.

\bibitem[\protect\citeauthoryear{Ansari, Stella, Turkmen, Zhang, Mercado, Shen,
  Shchur, Rangapuram, Pineda~Arango, Kapoor, Zschiegner, Maddix, Wang, Mahoney,
  Torkkola, Wilson, Bohlke-Schneider, and Wang}{Ansari
  et~al.}{2024}]{ansari2024chronos}
Ansari, A.~F., L.~Stella, C.~Turkmen, X.~Zhang, P.~Mercado, H.~Shen, O.~Shchur,
  S.~S. Rangapuram, S.~Pineda~Arango, S.~Kapoor, J.~Zschiegner, D.~C. Maddix,
  H.~Wang, M.~W. Mahoney, K.~Torkkola, A.~G. Wilson, M.~Bohlke-Schneider, and
  Y.~Wang (2024).
\newblock Chronos: Learning the language of time series.
\newblock {\em Transactions on Machine Learning Research\/}.

\bibitem[\protect\citeauthoryear{Basellini, Camarda, and Booth}{Basellini
  et~al.}{2023}]{basellini2023thirty}
Basellini, U., C.~G. Camarda, and H.~Booth (2023).
\newblock Thirty years on: A review of the {Lee--Carter} method for forecasting
  mortality.
\newblock {\em International Journal of Forecasting\/}~{\em 39\/}(3),
  1033--1049.

\bibitem[\protect\citeauthoryear{Basten, Sobotka, and Zeman}{Basten
  et~al.}{2014}]{basten2014future}
Basten, S., T.~Sobotka, and K.~Zeman (2014).
\newblock Future fertility in low fertility countries.
\newblock In W.~Lutz, W.~P. Butz, and S.~KC (Eds.), {\em World Population and
  Human Capital in the Twenty-First Century}, pp.\  39--146. Oxford University
  Press.

\bibitem[\protect\citeauthoryear{Bohk-Ewald, Li, and Myrskyl{\"a}}{Bohk-Ewald
  et~al.}{2018}]{bohkewald2018forecast}
Bohk-Ewald, C., P.~Li, and M.~Myrskyl{\"a} (2018).
\newblock Forecast accuracy hardly improves with method complexity when
  completing cohort fertility.
\newblock {\em Proceedings of the National Academy of Sciences\/}~{\em
  115\/}(37), 9187--9192.

\bibitem[\protect\citeauthoryear{Bongaarts}{Bongaarts}{2006}]{bongaarts2006causes}
Bongaarts, J. (2006).
\newblock The causes of stalling fertility transitions.
\newblock {\em Studies in Family Planning\/}~{\em 37\/}(1), 1--16.

\bibitem[\protect\citeauthoryear{Bongaarts and Sobotka}{Bongaarts and
  Sobotka}{2012}]{bongaarts2012demographic}
Bongaarts, J. and T.~Sobotka (2012).
\newblock A demographic explanation for the recent rise in {European}
  fertility.
\newblock {\em Population and Development Review\/}~{\em 38\/}(1), 83--120.

\bibitem[\protect\citeauthoryear{Camarda}{Camarda}{2019}]{camarda2019smooth}
Camarda, C.~G. (2019).
\newblock Smooth constrained mortality forecasting.
\newblock {\em Demographic Research\/}~{\em 41\/}(38), 1091--1130.

\bibitem[\protect\citeauthoryear{Camarda and Basellini}{Camarda and
  Basellini}{2021}]{camarda2021smoothing}
Camarda, C.~G. and U.~Basellini (2021).
\newblock Smoothing, decomposing and forecasting mortality rates.
\newblock {\em European Journal of Population\/}~{\em 37\/}(3), 569--602.

\bibitem[\protect\citeauthoryear{Cho, van Merri{\"e}nboer, G{\"u}l{\c{c}}ehre,
  Bahdanau, Bougares, Schwenk, and Bengio}{Cho et~al.}{2014}]{cho2014learning}
Cho, K., B.~van Merri{\"e}nboer, {\c{C}}.~G{\"u}l{\c{c}}ehre, D.~Bahdanau,
  F.~Bougares, H.~Schwenk, and Y.~Bengio (2014).
\newblock Learning phrase representations using {RNN} encoder--decoder for
  statistical machine translation.
\newblock In {\em Proceedings of the 2014 Conference on Empirical Methods in
  Natural Language Processing (EMNLP)}, pp.\  1724--1734.

\bibitem[\protect\citeauthoryear{Craven and Shavlik}{Craven and
  Shavlik}{1996}]{craven1996extracting}
Craven, M.~W. and J.~W. Shavlik (1996).
\newblock Extracting tree-structured representations of trained networks.
\newblock In D.~S. Touretzky, M.~C. Mozer, and M.~E. Hasselmo (Eds.), {\em
  Advances in Neural Information Processing Systems 8}, Cambridge, MA, pp.\
  24--30. MIT Press.

\bibitem[\protect\citeauthoryear{Currie, Durban, and Eilers}{Currie
  et~al.}{2004}]{currie2004smoothing}
Currie, I.~D., M.~Durban, and P.~H.~C. Eilers (2004).
\newblock Smoothing and forecasting mortality rates.
\newblock {\em Statistical Modelling\/}~{\em 4\/}(4), 279--298.

\bibitem[\protect\citeauthoryear{Ellison, Bijak, and Dodd}{Ellison
  et~al.}{2026}]{ellison2026parity}
Ellison, J., J.~Bijak, and E.~Dodd (2026).
\newblock Can incorporating parity information improve the reliability of
  completed cohort fertility projections? {Insights} from a {Bayesian}
  generalized additive model approach.
\newblock {\em Demography\/}~{\em 63\/}(2), 453--481.

\bibitem[\protect\citeauthoryear{Ellison, Dodd, and Forster}{Ellison
  et~al.}{2020}]{ellison2020forecasting}
Ellison, J., E.~Dodd, and J.~J. Forster (2020).
\newblock Forecasting of cohort fertility under a hierarchical {Bayesian}
  approach.
\newblock {\em Journal of the Royal Statistical Society Series A: Statistics in
  Society\/}~{\em 183\/}(3), 829--856.

\bibitem[\protect\citeauthoryear{{GBD 2021 Fertility and Forecasting
  Collaborators}}{{GBD 2021 Fertility and Forecasting
  Collaborators}}{2024}]{gbd2024global}
{GBD 2021 Fertility and Forecasting Collaborators} (2024).
\newblock Global fertility in 204 countries and territories, 1950--2021, with
  forecasts to 2100: a comprehensive demographic analysis for the {Global
  Burden of Disease Study} 2021.
\newblock {\em The Lancet\/}~{\em 403\/}(10440), 2057--2099.

\bibitem[\protect\citeauthoryear{Gietel-Basten and Sobotka}{Gietel-Basten and
  Sobotka}{2020}]{gietel2020uncertain}
Gietel-Basten, S. and T.~Sobotka (2020).
\newblock Uncertain population futures: Critical reflections on the {IHME}
  scenarios of future fertility, mortality, migration and population trends
  from 2017 to 2100.
\newblock SocArXiv.

\bibitem[\protect\citeauthoryear{Gietel-Basten and Sobotka}{Gietel-Basten and
  Sobotka}{2021}]{gietel2021trends}
Gietel-Basten, S. and T.~Sobotka (2021).
\newblock Trends in population health and demography.
\newblock {\em The Lancet\/}~{\em 398\/}(10300), 580--581.

\bibitem[\protect\citeauthoryear{Goldstein, Sobotka, and
  Jasilioniene}{Goldstein et~al.}{2009}]{goldstein2009end}
Goldstein, J.~R., T.~Sobotka, and A.~Jasilioniene (2009).
\newblock The end of ``lowest-low'' fertility?
\newblock {\em Population and Development Review\/}~{\em 35\/}(4), 663--699.

\bibitem[\protect\citeauthoryear{Goujon}{Goujon}{2025}]{goujon2025}
Goujon, A. (2025).
\newblock The demographic future that we do not know about.
\newblock {\em Science\/}~{\em 390\/}(6775), eaed1652.

\bibitem[\protect\citeauthoryear{Hewamalage, Bergmeir, and Bandara}{Hewamalage
  et~al.}{2021}]{hewamalage2021recurrent}
Hewamalage, H., C.~Bergmeir, and K.~Bandara (2021).
\newblock Recurrent neural networks for time series forecasting: Current status
  and future directions.
\newblock {\em International Journal of Forecasting\/}~{\em 37\/}(1), 388--427.

\bibitem[\protect\citeauthoryear{Hyndman and Ullah}{Hyndman and
  Ullah}{2007}]{hyndman2007robust}
Hyndman, R.~J. and M.~S. Ullah (2007).
\newblock Robust forecasting of mortality and fertility rates: A functional
  data approach.
\newblock {\em Computational Statistics \& Data Analysis\/}~{\em 51\/}(10),
  4942--4956.

\bibitem[\protect\citeauthoryear{Januschowski, Gasthaus, Wang, Salinas,
  Flunkert, Bohlke-Schneider, and Callot}{Januschowski
  et~al.}{2020}]{januschowski2020criteria}
Januschowski, T., J.~Gasthaus, Y.~Wang, D.~Salinas, V.~Flunkert,
  M.~Bohlke-Schneider, and L.~Callot (2020).
\newblock Criteria for classifying forecasting methods.
\newblock {\em International Journal of Forecasting\/}~{\em 36\/}(1), 167--177.

\bibitem[\protect\citeauthoryear{Jones}{Jones}{2007}]{jones2007delayed}
Jones, G.~W. (2007).
\newblock Delayed marriage and very low fertility in {Pacific Asia}.
\newblock {\em Population and Development Review\/}~{\em 33\/}(3), 453--478.

\bibitem[\protect\citeauthoryear{KC, Dhakad, Potan{\v{c}}okov{\'a}, Adhikari,
  Yildiz, Mamolo, Sobotka, Zeman, Abel, Lutz, and Goujon}{KC
  et~al.}{2024}]{kc2024updating}
KC, S., M.~Dhakad, M.~Potan{\v{c}}okov{\'a}, S.~Adhikari, D.~Yildiz, M.~Mamolo,
  T.~Sobotka, K.~Zeman, G.~Abel, W.~Lutz, and A.~Goujon (2024).
\newblock Updating the {Shared Socioeconomic Pathways} ({SSPs}) global
  population and human capital projections.
\newblock IIASA Working Paper WP-24-003, International Institute for Applied
  Systems Analysis, Laxenburg, Austria.

\bibitem[\protect\citeauthoryear{Koenker and Bassett}{Koenker and
  Bassett}{1978}]{koenker1978regression}
Koenker, R. and G.~Bassett, Jr. (1978).
\newblock Regression quantiles.
\newblock {\em Econometrica\/}~{\em 46\/}(1), 33--50.

\bibitem[\protect\citeauthoryear{Kohler, Billari, and Ortega}{Kohler
  et~al.}{2002}]{kohler2002emergence}
Kohler, H.-P., F.~C. Billari, and J.~A. Ortega (2002).
\newblock The emergence of lowest-low fertility in {Europe} during the 1990s.
\newblock {\em Population and Development Review\/}~{\em 28\/}(4), 641--680.

\bibitem[\protect\citeauthoryear{Lee and Carter}{Lee and
  Carter}{1992}]{lee1992modeling}
Lee, R.~D. and L.~R. Carter (1992).
\newblock Modeling and forecasting {U.S.} mortality.
\newblock {\em Journal of the American Statistical Association\/}~{\em
  87\/}(419), 659--671.

\bibitem[\protect\citeauthoryear{Li and Lee}{Li and Lee}{2005}]{li2005coherent}
Li, N. and R.~Lee (2005).
\newblock Coherent mortality forecasts for a group of populations: An extension
  of the {Lee--Carter} method.
\newblock {\em Demography\/}~{\em 42\/}(3), 575--594.

\bibitem[\protect\citeauthoryear{Lutz, Butz, and KC}{Lutz
  et~al.}{2014}]{lutz2014world}
Lutz, W., W.~P. Butz, and S.~KC (Eds.) (2014).
\newblock {\em World Population and Human Capital in the Twenty-First Century}.
\newblock Oxford: Oxford University Press.

\bibitem[\protect\citeauthoryear{Lutz, Goujon, KC, Stonawski, and
  Stilianakis}{Lutz et~al.}{2018}]{lutz2018demographic}
Lutz, W., A.~Goujon, S.~KC, M.~Stonawski, and N.~Stilianakis (2018).
\newblock {\em Demographic and human capital scenarios for the 21st century:
  2018 assessment for 201 countries}.
\newblock Luxembourg: Publications Office of the European Union.

\bibitem[\protect\citeauthoryear{Makridakis, Spiliotis, and
  Assimakopoulos}{Makridakis et~al.}{2022}]{makridakis2022m5}
Makridakis, S., E.~Spiliotis, and V.~Assimakopoulos (2022).
\newblock {M5} accuracy competition: Results, findings, and conclusions.
\newblock {\em International Journal of Forecasting\/}~{\em 38\/}(4),
  1346--1364.

\bibitem[\protect\citeauthoryear{Montero-Manso and Hyndman}{Montero-Manso and
  Hyndman}{2021}]{montero2021principles}
Montero-Manso, P. and R.~J. Hyndman (2021).
\newblock Principles and algorithms for forecasting groups of time series:
  Locality and globality.
\newblock {\em International Journal of Forecasting\/}~{\em 37\/}(4),
  1632--1653.

\bibitem[\protect\citeauthoryear{Park}{Park}{2025}]{park2025next}
Park, P. (2025).
\newblock Next-generation mortality forecasting with deep learning.
\newblock SocArXiv preprint.

\bibitem[\protect\citeauthoryear{Raftery, Alkema, and Gerland}{Raftery
  et~al.}{2014}]{raftery2014bayesian}
Raftery, A.~E., L.~Alkema, and P.~Gerland (2014).
\newblock Bayesian population projections for the {United Nations}.
\newblock {\em Statistical Science\/}~{\em 29\/}(1), 58--68.

\bibitem[\protect\citeauthoryear{Raftery and
  {\v{S}}ev{\v{c}}{\'\i}kov{\'a}}{Raftery and
  {\v{S}}ev{\v{c}}{\'\i}kov{\'a}}{2023}]{raftery2023probabilistic}
Raftery, A.~E. and H.~{\v{S}}ev{\v{c}}{\'\i}kov{\'a} (2023).
\newblock Probabilistic population forecasting: Short to very long-term.
\newblock {\em International Journal of Forecasting\/}~{\em 39\/}(1), 73--97.

\bibitem[\protect\citeauthoryear{Rudin}{Rudin}{2019}]{rudin2019stop}
Rudin, C. (2019).
\newblock Stop explaining black box machine learning models for high stakes
  decisions and use interpretable models instead.
\newblock {\em Nature Machine Intelligence\/}~{\em 1\/}(5), 206--215.

\bibitem[\protect\citeauthoryear{Salinas, Flunkert, Gasthaus, and
  Januschowski}{Salinas et~al.}{2020}]{salinas2020deepar}
Salinas, D., V.~Flunkert, J.~Gasthaus, and T.~Januschowski (2020).
\newblock {DeepAR}: Probabilistic forecasting with autoregressive recurrent
  networks.
\newblock {\em International Journal of Forecasting\/}~{\em 36\/}(3),
  1181--1191.

\bibitem[\protect\citeauthoryear{Schmertmann, Zagheni, Goldstein, and
  Myrskyl{\"a}}{Schmertmann et~al.}{2014}]{schmertmann2014bayesian}
Schmertmann, C., E.~Zagheni, J.~R. Goldstein, and M.~Myrskyl{\"a} (2014).
\newblock {Bayesian} forecasting of cohort fertility.
\newblock {\em Journal of the American Statistical Association\/}~{\em
  109\/}(506), 500--513.

\bibitem[\protect\citeauthoryear{Susmann and Alkema}{Susmann and
  Alkema}{2025}]{susmann2025flexible}
Susmann, H. and L.~Alkema (2025).
\newblock Flexible modelling of demographic transition processes with a
  {Bayesian} hierarchical {B}-splines model.
\newblock {\em Journal of the Royal Statistical Society Series C: Applied
  Statistics\/}~{\em 74\/}(5), 1340--1371.

\bibitem[\protect\citeauthoryear{Sutskever, Vinyals, and Le}{Sutskever
  et~al.}{2014}]{sutskever2014sequence}
Sutskever, I., O.~Vinyals, and Q.~V. Le (2014).
\newblock Sequence to sequence learning with neural networks.
\newblock In {\em Advances in Neural Information Processing Systems},
  Volume~27.

\bibitem[\protect\citeauthoryear{{United Nations}}{{United
  Nations}}{2024}]{un2024wpp}
{United Nations} (2024).
\newblock {\em World Population Prospects 2024: Summary of Results}.
\newblock New York: United Nations Department of Economic and Social Affairs,
  Population Division.

\bibitem[\protect\citeauthoryear{{United Nations}}{{United
  Nations}}{2025}]{un2025fertility}
{United Nations} (2025).
\newblock {\em World Fertility 2024}.
\newblock New York: United Nations Department of Economic and Social Affairs,
  Population Division.
\newblock UN DESA/POP/2024/TR/NO.11.

\bibitem[\protect\citeauthoryear{Wen, Torkkola, Narayanaswamy, and Madeka}{Wen
  et~al.}{2017}]{wen2017multi}
Wen, R., K.~Torkkola, B.~Narayanaswamy, and D.~Madeka (2017).
\newblock A multi-horizon quantile recurrent forecaster.
\newblock arXiv preprint arXiv:1711.11053.

\end{thebibliography}

\end{document}